\documentclass[preprint,aps,superscriptaddress,showpacs,nofootinbib,eqsecnum,preprintnumbers,amsmath,amssymb,tightenlines]{revtex4-1}
\usepackage{graphicx}
\usepackage{dcolumn}
\usepackage{bm}
\usepackage{amsfonts,amssymb,amsmath,amsthm,amsxtra,mathtools,euscript,mathrsfs}
\usepackage{commath}
\usepackage{MnSymbol}
\usepackage{amsthm}
\usepackage{hyperref}
\usepackage[makeroom]{cancel}
\usepackage{subfigure}
\newtheorem*{theorem*}{Theorem}
\def\be{\begin{eqnarray}}
\def\ee{\end{eqnarray}}

\begin{document}     

\title{On the Dynamical Instability of Monatomic Fluid Spheres in (N+1)-dimensional Spacetime}

\author{Wei-Xiang Feng}
\email{wfeng016@ucr.edu}
\affiliation{Department of Physics and Astronomy, University of California, Riverside, CA 92521, USA}



\begin{abstract}
In this note, I derive the Chandrasekhar instability of a fluid sphere in ($N$+1)-dimensional Schwarzschild-Tangherlini spacetime and take the homogeneous (uniform energy density) solution for illustration. Qualitatively, the effect of positive (negative) cosmological constant tends to destabilize (stabilize) the sphere. In the absence of cosmological constant, the privileged position of (3+1)-dimensional spacetime is manifest in its own right. As it is, the \emph{marginal dimensionality} in which a monatomic ideal fluid sphere is \emph{stable but not too stable} to trigger the onset of gravitational collapse. Furthermore, it is 
the \emph{unique} dimensionality that can accommodate stable hydrostatic equilibrium with a \emph{positive cosmological constant.} However, given the current cosmological constant observed, no stable configuration can be larger than $10^{21}~{\rm M}_\odot$.  On the other hand, in (2+1) dimensions, it is \emph{too stable} either in the context of Newtonian Gravity (NG) or Einstein's General Relativity (GR). In GR, the role of \emph{negative cosmological constant} is crucial not only to guarantee fluid equilibrium (decreasing monotonicity of pressure) but also to have the Ba{\~n}ados-Teitelboim-Zanelli (BTZ) black hole solution. Owing to the negativeness of the cosmological constant, there is no unstable configuration for a homogeneous fluid disk with mass $0<\mathcal{M}\leq0.5$ to collapse into a naked singularity, which supports the Cosmic Censorship Conjecture. However, the relativistic instability can be triggered for a homogeneous disk with mass $0.5<\mathcal{M}\lesssim0.518$ under causal limit, which implies that BTZ holes of mass $\mathcal{M}_{\rm BTZ}>0$ could emerge from collapsing fluid disks under proper conditions. 
The implicit assumptions and implications are also discussed.
\end{abstract}

                                               
\maketitle


\section{\label{sec:intro}introduction}
Dimensionality of spacetime is a question with a long history~\cite{Ehrenfest:1918,Ehrenfest:1920,Whitrow:1955,Tangherlini:1963bw,Barrow:1983,Barrow:1988yia,Caruso:1986de,Tegmark:1997jg}. Starting from
Ehrenfest~\cite{Ehrenfest:1918,Ehrenfest:1920}, who argued based on the ``stability postulate'' of the two-body problem, the fundamental laws of physics favor (3+1) dimensions. His approach is still valid in the framework of general relativity (GR) as well as hydrogen atom, as was shown by Tangherlini~\cite{Tangherlini:1963bw}, however, see Ref.~\cite{Caruso:1986de} for an alternative procedure and references therein.
Tegmark~\cite{Tegmark:1997jg} argued that the existence of only one temporal dimension by requiring hyperbolic equations of motion, and hence \emph{predictability}, leaves the question why three (macroscopic) spatial dimensions are favorable. Various arguments were proposed for the reasoning~\cite{Momen:2011jc,Gonzalez-Ayala:2015xda,Brandenberger:1988aj,Greene:2012sa,Durrer:2005nz,Nielsen:1993fd,Deser:2019oqc}, although the dimensionality could be dynamical and scale-dependent, as a physical observable~\cite{Carlip:2017eud}.


Presumably, the ``non-compact'' (3+1) dimensions might be just an illusion due to our human perception. However, constraint from gravitational waves indicates we certainly live in a universe of non-compact (3+1) dimensions~\cite{Pardo:2018ipy}, as described by GR.
On the other hand, although it was usually cited that structures in (2+1) are not complex enough to accommodate life, as opposed to common view, life could exist in (2+1) was discussed more recently~\cite{Scargill:2019rse}.

Stellar equilibrium in different dimensions also has been explored previously~\cite{PoncedeLeon:2000pj,Paul:2004nh,Zarro:2009gd}. One important feature on stellar stability in GR is the Buchdahl stability bound~\cite{Buchdahl:1959zz}. It states that the mass of a spherical compact object must exceed the $9/4$ of its gravitational radius in (3+1) dimensions. Otherwise, there is no stable stellar equilibrium, and  it would definitely collapse into a black hole. Buchdahl bound in higher dimensions~\cite{PoncedeLeon:2000pj} with cosmological constant~\cite{Zarro:2009gd} and its universality in other gravity theories~\cite{Goswami:2015dma,Feng:2018jrh,Chakraborty:2022jgl} were also investigated previously.
Nonetheless, the dynamical instability of a star might have set in well before the Buchdahl bound.

The dynamical instability of a self-gravitating sphere in the context of GR has been explored long ago by Chandrasekhar (1964)~\cite{Chandrasekhar:1964zza,Chandrasekhar:1964zz} and Zel'dovich \& Podurets (1966)~\cite{Zeldovich:1966}, via the study of pulsation equation and binding energy, respectively. It was found that the turning point of fractional binding energy~\cite{Ipser:1980,Gunther:2021pbf} is very close to the result using the pulsation equation~\cite{Feng:2021rst}. 
Chandrasekhar's criteron~\cite{Chandrasekhar:1964zza,Chandrasekhar:1964zz} provides a sufficient condition triggering the black hole formation.
In hydrostatic equilibrium, the pressure of a star balances its self-gravity. As a gravitationally bound system, a star behaves as if it has negative specific heat: The more energy it loses, the hotter it becomes~\cite{LyndenBell:1968yw,Spitzer:1987aa}. Therefore it undergoes the gravothermal evolution due to the heat dissipations. At each evolution stage the instability might set in depending on the fluid's \emph{stiffness}, which is characterized by its adiabatic index~\cite{Feng:2021rst}. 
For example, the adiabatic index of a (3+1) monatomic ideal fluid  transitions from $5/3$ (stiff) toward $4/3$ (soft) when the particles become more relativistic through the conversion of gravitational energy.
This method has also been extended to address the stellar instability in (3+1) spacetime with a non-zero cosmological constant~\cite{Boehmer:2005kk,Posada:2020svn}, the extra dimension influence~\cite{Arbanil:2019mae} on the strange quark stars, the stability of supermassive stars~\cite{Haemmerle:2020wwz}, and the self-interacting dark halo core collapse~\cite{Feng:2020kxv,Feng:2021rst}.
However, see also the gravothermal instabilities from energy consideration in~\cite{Roupas:2014sda,Roupas:2018zie,Roupas:2020tcn}.

In (2+1) spacetime, a negative cosmological constant is to guarantee not only hydrostatic equilibrium~\cite{Cruz:1994ar} (the pressure is monotonically decreasing) but also permit a black hole solution of Ba{\~n}ados-Teitelboim-Zanelli (BTZ)~\cite{Banados:1992wn}.
The dynamical process of dust collapse~\cite{Ross:1992ba}, critical collapse of scalar field~\cite{Pretorius:2000yu,Husain:2000vm,Jalmuzna:2015hoa}, and ultra-relativistic fluid~\cite{Bourg:2021ewu,Bourg:2021vpv} into a BTZ hole have been shown to be possible.
Nevertheless, in this note, we are more interested in the condition triggering the dynamical instability of a monatomic fluid disk in the \emph{hydrodynamic} limit.
The pressure in a fluid can not be ignored because it could prevent the fluid from further collapse.
On the other hand, as the random motion of the particles in the fluid influences the gravitational potential in the macroscopic picture through the pressure effect, (2+1) static stars of perfect fluid qualitatively differ in their behavior from that of dust~\cite{Giddings:1983es}.
As a result, rather than counteract the gravitational attraction, it could further destabilize the fluid disk at some point, just as the cases of (3+1). However, the negative cosmological constant introduced tends to stabilize the fluid. Thus, the competition between the pressure (relativistic) effect and the cosmological constant is crucial to trigger the black hole formation.

In this work, we examine the \emph{space dimensionality} $N$ from the viewpoint of the Chandrasekhar instability of an ideal monatomic fluid sphere in ($N$+1) dimensions. The paper is organized as follows: We first derive the ($N$+1) pulsation equation of a perfect fluid sphere with and without cosmological constant and the corresponding Chandrasekhar's criterion in Sec.~\ref{sec:schw}. As an illustration, we then present the ($N$+1) homogeneous (uniform energy density) solution in Sec.~\ref{sec:homogeneous} and numerically determine the condition at the onset of instabilities in Sec.~\ref{sec:numerics}. We briefly summarize the results with the implicit assumptions, and discuss the physical implications in Sec.~\ref{sec:discuss}. 
In particular, we assume GR equations hold, the fluid sphere is homogeneous and monatomic.
Geometric unit ($G_N=c=1$) is used throughout the text, where $G_N$ is the Newton's constant in ($N$+1) dimensions.

\section{\label{sec:schw}(N+1)-dimensional spacetime of spherical symmetry}
We consider a spherically symmetric spacetime in ($N$+1) dimensions, 
\begin{equation}\label{seceq:metric1}
{\rm d}s^2=-e^{2\Phi(t,r)}{\rm d}t^2+e^{2\Lambda(t,r)}{\rm d}r^2+r^2{\rm d}\Omega_{N-1}^2,
\end{equation}
where 
\begin{equation}\label{seceq:metric2}
{\rm d}\Omega^{2}_{N-1}={\rm d}\theta_{1}^{2}+\sin^2\theta_{1}{\rm d}\theta_{2}^2+...+\prod_{j=1}^{N-2}\sin^2\theta_{j}{\rm d}\theta_{N-1}^2.
\end{equation}

After the standard calculations, the field equations $G^\mu{}_\nu=\kappa_N T^\mu{}_\nu$ give
\begin{subequations}
\begin{equation}\label{seceq:tt}
-\frac{(N-1)\Lambda'}{r}e^{-2\Lambda}-\frac{(N-1)(N-2)}{2r^2}\left[1-e^{-2\Lambda}\right]=\kappa_{N} T^{t}{}_{t},
\end{equation}
\begin{equation}\label{seceq:rr}
\frac{(N-1)\Phi{'}}{r}e^{-2\Lambda}-\frac{(N-1)(N-2)}{2r^2}\left[1-e^{-2\Lambda}\right]=\kappa_{N} T^{r}{}_{r},
\end{equation}
\begin{equation}\label{seceq:rt}
\frac{N-1}{r}\dot{\Lambda}e^{-2\Lambda}=\kappa_{N} T^{r}{}_{t},
\end{equation}
\begin{equation}\label{seceq:theta}
\begin{aligned}
&\left[\Phi{''}+\Phi{'}^2-\Phi{'}\Lambda{'}+\frac{(N-2)}{r}(\Phi{'}-\Lambda{'})\right]e^{-2\Lambda}\\
&-\frac{(N-2)(N-3)}{2r^2}\left[1-e^{-2\Lambda}\right]-(\ddot{\Lambda}+\dot{\Lambda}^2-\dot{\Phi}\dot{\Lambda})e^{-2\Phi}=\kappa_NT^{\theta_{1}}{}_{\theta_{1}},
\end{aligned}
\end{equation}
\end{subequations}
where the ``prime'' and ``dot'' denote the ``radial'' and ``time'' derivatives, respectively, and $T^{\theta_{1}}{}_{\theta_{1}}=T^{\theta_{2}}{}_{\theta_{2}}=...=T^{\theta_{N-1}}{}_{\theta_{N-1}}\equiv p$ because of spherical symmetry. On the other hand, combining Eq.~(\ref{seceq:tt}) and (\ref{seceq:rr}), we obtain 
\begin{equation}\label{seceq:ttrr}
\frac{(N-1)}{r}(\Phi{'}+\Lambda{'})e^{-2\Lambda}=\kappa_{N}(T^{r}{}_{r}-T^{t}{}_{t}).
\end{equation}

The conservation of energy-momentum tensor $\nabla_\mu T^\mu{}_\nu=0$ leads to
\begin{subequations}
\begin{equation}\label{seceq:conserv1}
\partial_t T^{t}{}_{t}+\partial_r T^{r}{}_{t}+(T^{t}{}_{t}-T^{r}{}_{r})\dot{\Lambda}+T^{r}{}_{t}\left(\Phi{'}+\Lambda{'}+\frac{N-1}{r}\right)=0
\end{equation}
and
\begin{equation}\label{seceq:conserv2}
\partial_t T^{t}{}_{r}+\partial_r T^{r}{}_{r}+T^{t}{}_{r}(\dot{\Phi}+\dot{\Lambda})+(T^{r}{}_{r}-T^{t}{}_{t})\Phi{'}+\frac{N-1}{r}(T^{r}{}_{r}-p)=0.
\end{equation}
\end{subequations}

If we define the Schwarzschild mass function $M(r)$ through 
\begin{equation}\label{seceq:mass1}
1-e^{-2\Lambda}\equiv 2M/r^{N-2},
\end{equation}
then $G^{t}{}_{t}=\kappa_{N} T^{t}{}_{t}=-\kappa_{N}\rho$ leads to $M'(r)=[\kappa_{N}/(N-1)]\rho r^{N-1}$. Accordingly, the mass function
\begin{equation}\label{seceq:mass2}
M(r)=\omega_N\int_0^r \rho \bar{r}^{N-1}d\bar{r},
\end{equation}
where
\begin{equation}\label{seceq:area}
\omega_{N}=\frac{2\pi^{N/2}}{\Gamma(N/2)}=\frac{\kappa_N}{N-1}
\end{equation}
is the area of the unit sphere in $N$-dimensional space. Therefore, by construction, the Einstein coupling constant $\kappa_{N}=(N-1)\omega_{N}$ automatically imposes the vacuum identity 
$G^\mu{}_\nu=0$ of Einsteinian gravity in (1+1) dimensions.

Now we consider isotropic pressure ($T^r{}_r=p$) in static situation, Eq.~(\ref{seceq:conserv2}) leads to $p'=-(\rho+p)\Phi'$, and $\Phi'$ can be replaced via Eq.~(\ref{seceq:rr}), resulting in the Tolman-Oppenheimer-Volkoff (TOV) equation in ($N$+1) dimensions:
\begin{equation}\label{seceq:tov}
\left(1-\frac{2M}{r^{N-2}}\right)p'=-(\rho+p)\left(\frac{(N-2)M}{r^{N-1}}+\frac{\kappa_N}{N-1} pr\right).
\end{equation}

Equation-of-state (EoS) $p=p(\rho)$ and boundary conditions $M(0)=0$, $p(R)=0$ must be imposed in order to determine the total fluid mass $\mathcal{M}\equiv M(R)$.

\subsection{\label{sec:linearperturb}Linear radial perturbation \& the adiabatic index}
The perfect fluid description $T^\mu{}_\nu=(\rho+p)u^\mu u_\nu+p\delta^\mu{}_\nu$ implies $T^r{}_t=0$ in the static case. However, under the \emph{linear radial perturbation} $\partial_t\xi=\dot{\xi}\equiv dr/dt$, 
\begin{equation}\label{seceq:prt}
T^{r}{}_{t}=-(\rho+p)\dot{\xi},
\end{equation}
where $u^\mu$ is the 4-velocity of the fluid element, and we have introduced the ``Lagrangian displacement'' $\xi$ as well as $u_t=-e^\Phi$, $u^r=e^{-\Phi}\dot{\xi}$.

Now we denote all the variables $X$'s in equilibrium  by $X_0(r)$, which is independent of time.  After perturbation $X(t, r)=X_0(r)+\delta X(t, r)$, only the perturbed quantities have the time dependence, where $\delta$ denotes the ``Eulerian change'' of the perturbation. Keeping only the terms of \emph{first-order corrections} from Eqs.~(\ref{seceq:tt}), (\ref{seceq:rr}), (\ref{seceq:rt}), (\ref{seceq:conserv1}) and (\ref{seceq:conserv2}), we have the linearized equations governing the perturbation
\begin{subequations}
\begin{equation}\label{seceq:ptt}
\left[r^{N-2}e^{-2\Lambda_{0}}2\delta\Lambda\right]'=\frac{2\kappa_N}{N-1} r^{N-1}\delta\rho,
\end{equation}
\begin{equation}\label{seceq:prr}
\frac{N-1}{r}e^{-2\Lambda_{0}}(\delta\Phi{'}-2\Phi_{0}{'}\delta\Lambda)-\frac{(N-1)(N-2)}{r^2}e^{-2\Lambda_{0}}\delta\Lambda=\kappa_N \delta p,
\end{equation}
\begin{equation}\label{seceq:prt}
\frac{N-1}{r}e^{-2\Lambda_{0}}\delta\dot{\Lambda}=-\kappa_N (\rho_{0}+p_{0})\dot{\xi}
=-\frac{N-1}{r}e^{-2\Lambda_{0}}(\Phi_{0}{'}+\Lambda_{0}{'})\dot{\xi},
\end{equation}
\begin{equation}\label{seceq:pconserv1}
\delta\dot{\rho}+\left[(\rho_0+p_0)\dot{\xi}\right]'+(\rho_0+p_0)\left[\delta\dot{\Lambda}+\dot{\xi}\left(\Phi_0'+\Lambda_0'+\frac{N-1}{r}\right)\right]=0,
\end{equation}
\begin{equation}\label{seceq:pconserv2}
e^{2(\Lambda_{0}-\Phi_{0})}(\rho_{0}+p_{0})\ddot{\xi}+\delta p'+(\rho_{0}+p_{0})\delta\Phi{'}+(\delta\rho+\delta p)\Phi_{0}{'}=0,
\end{equation}
\end{subequations}
respectively. 
In addition, Eq.~(\ref{seceq:prt}) also leads to
\begin{equation}\label{seceq:prtextra}
\delta\Lambda=-\xi(\Phi_{0}{'}+\Lambda_{0}{'}),
\end{equation}
thus Eq.~(\ref{seceq:ptt}) and (\ref{seceq:pconserv1}) (after performing time integration) identically lead to
\begin{subequations}\label{seceq:varrho}
\begin{equation}
\delta\rho=-\frac{1}{r^{N-1}}\left[r^{N-1}(\rho_{0}+p_{0})\xi\right]'=-\xi\frac{{\rm d}\rho_{0}}{{\rm d}r}-\xi\frac{{\rm d}p_{0}}{{\rm d}r}-(\rho_{0}+p_{0})\frac{1}{r^{N-1}}\left(r^{N-1}\xi\right)'
\end{equation}
or with the substitution $p_0'=-(\rho_{0}+p_{0})\Phi_{0}{'}$, we obtain
\begin{equation}
\Delta\rho=\delta\rho+\xi\frac{{\rm d}\rho_{0}}{{\rm d}r}=-(\rho_{0}+p_{0})\frac{e^{\Phi_{0}}}{r^{N-1}}\left(r^{N-1}e^{-\Phi_{0}}\xi\right)',
\end{equation}
\end{subequations}
where $\Delta=\delta+\xi\partial_r$ denotes the ``Lagrangian change'' of the perturbation. 
To express $\delta{p}$ in terms of $\xi$, we assume the conservation of the baryon number $\nabla_{\alpha}(nu^{\alpha})=0$, that is
\begin{equation}\label{seceq:conserv3}
\partial_{t}(ne^{-\Phi})+\partial_{r}(n\dot{\xi}e^{-\Phi})+ne^{-\Phi}(\dot{\Phi}+\dot{\Lambda})+ne^{-\Phi}\dot{\xi}\left(\Phi{'}+\Lambda{'}+\frac{N-1}{r}\right)=0.
\end{equation}

Keeping only the first-order terms of perturbation, one obtains
\begin{equation}\label{seceq:pconserv3}
e^{-\Phi_{0}}\delta{\dot{n}}+\frac{1}{r^{N-1}}\left(n_{0}r^{N-1}\dot{\xi}e^{-\Phi_{0}}\right)'+n_{0}e^{-\Phi_{0}}\delta{\dot{\Lambda}}+n_{0}e^{-\Phi_{0}}\dot{\xi}(\Phi_{0}{'}+\Lambda_{0}{'})=0.
\end{equation}

After integration over time and using Eq.~(\ref{seceq:prtextra}), reduce to
\begin{subequations}\label{seceq:varnum}
\begin{equation}
\delta{n}=-\frac{e^{\Phi_{0}}}{r^{N-1}}\left(n_{0}r^{N-1}{\xi}e^{-\Phi_{0}}\right)'=-\xi\frac{{\rm d}n_{0}}{{\rm d}r}-n_{0}\frac{e^{\Phi_{0}}}{r^{N-1}}\left(r^{N-1}e^{-\Phi_{0}}\xi\right)'
\end{equation}
or
\begin{equation}
\Delta n\equiv\delta{n}+\xi\frac{{\rm d}n_{0}}{{\rm d}r}=-n_{0}\frac{e^{\Phi_{0}}}{r^{N-1}}\left(r^{N-1}e^{-\Phi_{0}}\xi\right)'.
\end{equation}
\end{subequations}

The adiabatic perturbation of the pressure is related to that of the number density through adiabatic index; if the EoS $n=n(\rho, p)$ is given, then
\[
\delta{n}=\left(\frac{\partial{n}}{\partial{\rho}}\right)_{p}\delta{\rho}+\left(\frac{\partial{n}}{\partial{p}}\right)_{\rho}\delta{p}
\]
together with Eq.~(\ref{seceq:varrho}) and (\ref{seceq:varnum}) lead to
\[
-\xi\frac{{\rm d}n_{0}}{{\rm d}r}-n_{0}\frac{e^{\Phi_{0}}}{r^{N-1}}\left(r^{N-1}e^{-\Phi_{0}}\xi\right)'=\left(\frac{\partial{n}}{\partial{\rho}}\right)_p\left[-\xi\frac{{\rm d}\rho_{0}}{{\rm d}r}-(\rho_{0}+p_{0})\frac{e^{\Phi_{0}}}{r^{N-1}}\left(r^{N-1}e^{-\Phi_{0}}\xi\right)'\right]+\left(\frac{\partial{n}}{\partial{p}}\right)_\rho\delta{p}.
\]

After the substitution
\[
\frac{{\rm d}n_{0}}{{\rm d}r}=\left(\frac{\partial{n}}{\partial\rho}\right)_p\frac{{\rm d}\rho_{0}}{{\rm d}r}+\left(\frac{\partial{n}}{\partial{p}}\right)_\rho\frac{{\rm d}p_{0}}{{\rm d}r},
\]
one finds
\begin{subequations}\label{seceq:varp}
\begin{equation}
\delta{p}=-\xi\frac{{\rm d}p_{0}}{{\rm d}r}-\gamma{p_{0}}\frac{e^{\Phi_{0}}}{r^{N-1}}\left(r^{N-1}e^{-\Phi_{0}}\xi\right)'
\end{equation}
or
\begin{equation}
\Delta p\equiv \delta{p}+\xi\frac{{\rm d}p_{0}}{{\rm d}r}=-\gamma{p_{0}}\frac{e^{\Phi_{0}}}{r^{N-1}}\left(r^{N-1}e^{-\Phi_{0}}\xi\right)',
\end{equation}
\end{subequations}
and the adiabatic index $\gamma$ of the fluid is defined by
\begin{subequations}\label{seceq:adiabindex}
\begin{equation}
\gamma\equiv\frac{1}{p_{0}\left(\partial{n}/\partial{p}\right)_{\rho}}\left[n_{0}-(\rho_{0}+p_{0})\left(\frac{\partial{n}}{\partial\rho}\right)_{p}\right]
=\frac{\rho_{0}+p_{0}}{p_{0}}\left(\frac{\partial{p}}{\partial\rho}\right)_s
=\frac{n_0}{p_0}\left(\frac{\partial p}{\partial n}\right)_s,
\end{equation}
where Eqs.~(\ref{appeq:thermo2}), (\ref{appeq:thermo3}) in App.~\ref{app:thermo} have been used to reach the last equality, and the subscript $s$ denotes that the change is adiabatic. This also implies the Lagrangian change is equivalent to the adiabatic change in view of Eqs.~(\ref{seceq:varrho}), (\ref{seceq:varnum}), (\ref{seceq:varp}), and one can write
\begin{equation}
\gamma=\left(\frac{\partial\ln p}{\partial\ln n}\right)_{s}\equiv\frac{\Delta\ln p}{\Delta\ln n},
\end{equation}
\end{subequations}
which is the generic definition and can be determined given a EoS.
In the literature, the symbol of adiabatic index is conventionally denoted as $\Gamma_1$ and $\gamma\equiv{c_p}/{c_n}$, where $c_p$, $c_n$ denote isobaric, isochoric specific heat capacities, respectively. The two indices are related by $\Gamma_1=\chi_T\gamma$, where $\chi_T\equiv\left({\partial\ln p}/{\partial\ln n}\right)_{T}$ evaluated at constant temperature $T$. However, for an ideal gas without radiation pressure, $\chi_T=1$ and so $\Gamma_1=\gamma$~\cite{Ogilvie:2016}.

\subsection{\label{sec:adiab}The adiabatic index of an ideal monatomic fluid}
For an ideal monatomic fluid, the adiabatic index depends on the \emph{degrees of freedom} of spatial dimensions. Given a distribution function $f(\mathbf{x}, \mathbf{p})$ of monatomic particles with phase space measure ${\rm d}^N\mathbf{x}~{\rm d}^N\mathbf{p}$, the EoS can be determined by
\begin{subequations}
\begin{equation}
n(\mathbf{x})=\int f(\mathbf{x}, \mathbf{p})~{\rm d}^{N}\mathbf{p},
\end{equation}
\begin{equation}
\rho(\mathbf{x})=\int E f(\mathbf{x}, \mathbf{p})~{\rm d}^{N}\mathbf{p},
\end{equation}
\begin{equation}
p(\mathbf{x})=\frac{1}{N}\int \mathbf{p}\cdot\frac{\partial E}{\partial\mathbf{p}}f(\mathbf{x}, \mathbf{p})~{\rm d}^{N}\mathbf{p}=\frac{1}{N}\int \frac{\mathbf{p}^2}{E}f(\mathbf{x}, \mathbf{p})~{\rm d}^{N}\mathbf{p},
\end{equation}
\end{subequations}
where $E=\sqrt{\mathbf{p}^2+m^2}$ the energy of the particle with rest mass $m$, and $N$ in the denominator of the pressure expression is due to \emph{equipartition theorem}.

The EoS can be prescribed by the $\gamma$-law form $p=K(mn)^{\gamma}$~\cite{Tooper:1965}, which satisfies the definition of adiabatic index provided that $K$, $\gamma$ are \emph{not explicit} functions of $n$ under adiabatic perturbation.
The first law of thermodynamics ${\rm d(internal~energy)}=-p{\rm~d(volume)}$, under adiabatic change and particle number conservation, results in
\begin{equation*}
{\rm d}\left(\frac{u}{n}\right)=-p~{\rm d}\left(\frac{1}{n}\right)=\frac{p}{n^2}{\rm d}n=Km^\gamma n^{\gamma-2} {\rm d}n.
\end{equation*}

Direct integration gives $u/n=Km^\gamma(\gamma-1)^{-1}n^{\gamma-1}$, thus the internal energy density
\begin{equation}
u\equiv\rho-mn=(\gamma-1)^{-1}p.
\end{equation}

It turns out that the adiabatic index
\begin{equation}
\gamma=1+\frac{p}{\rho-mn}
\end{equation}
depends on the spatial dimensions $N$ and the relativistic extent of the particles, specifically, in non-relativistic limit $\abs{\mathbf{p}}\ll m$, $\gamma\rightarrow 1+2/N$; ultra-relativistic limit $\abs{\mathbf{p}}\gg m$, $\gamma\rightarrow1+1/N$. Regardless of the distribution $f(\mathbf{x}, \mathbf{p})$ classical or quantum, the variation of $\gamma$ actually depends the velocity dispersion 
\begin{equation}\label{seceq:vdis}
v\equiv\sqrt{\frac{Np}{\rho}}
\end{equation} 
via (see App.~\ref{app:idealgas} for an explicit example)
\begin{equation}\label{seceq:adiabideal}
\gamma(v)=1+\frac{1+\sqrt{1-v^2}}{N}.
\end{equation} 
As an aside, we note that this applies only for ideal (classical or quantum) fluids. If the \emph{microscopic interaction} between particles is significant, the internal energy density $u$ will contain interacting energy between particles, and $K$, $\gamma$ might depend explicitly on $n$.

\subsection{\label{sec:puls}The pulsation equation and the critical adiabatic index}
Using Eqs.~(\ref{seceq:prr}), (\ref{seceq:prt}) and (\ref{seceq:prtextra}), one can derive
\[
(\rho_{0}+p_{0})\delta\Phi{'}=\left[\delta p-(\rho_{0}+p_{0})\left(2\Phi_{0}{'}+\frac{N-2}{r}\right)\xi\right](\Phi_{0}{'}+\Lambda_{0}{'}).
\]

Substitution of $\delta\Phi'$ for Eq.~(\ref{seceq:pconserv2}) and assuming all the perturbed quantities have the time-dependence of the form $e^{i\omega{t}}$ with the eigenfrequency $\omega$, one can show that 
\[
\omega^{2}e^{2(\Lambda-\Phi)}(\rho+p){\xi}=\delta p'+\delta p(2\Phi{'}+\Lambda{'})+\delta\rho\Phi{'}-(\rho+p)\left(2\Phi{'}+\frac{N-2}{r}\right)(\Phi{'}+\Lambda{'})\xi,
\]
in which we drop the subscript ``zero'' for simplicity hereafter.

Now we can further simplify the result by replacing the perturbed quantities (except $\xi$) with the unperturbed ones in equilibrium. With the proper substitutions via Eqs.~(\ref{seceq:theta}), (\ref{seceq:tov}), (\ref{seceq:varrho}), and (\ref{seceq:varp}), one can derive
the ``pulsation equation''
\begin{align}\label{seceq:puls1}
\nonumber
\omega^{2}e^{2(\Lambda-\Phi)}(\rho+p){\xi}=
&\frac{2(N-1)}{r}\frac{{\rm d}p}{{\rm d}r}\xi-e^{-(2\Phi+\Lambda)}\left[e^{3\Phi+\Lambda}\frac{\gamma p}{r^{N-1}}\left(r^{N-1}e^{-\Phi}\xi\right)'\right]'\\
&+\frac{2\kappa_N}{N-1} e^{2\Lambda}p(\rho+p)\xi-\frac{1}{\rho+p}\left(\frac{{\rm d}p}{{\rm d}r}\right)^2\xi,
\end{align}
governing the linear instability at the first-order with boundary conditions $\xi=0$ at $r=0$ and $\delta{p}=0$ at $r=R$ (radius of the sphere). Clearly, $N=3$ reduces to the result derived by Chandrasekhar (1964)~\cite{Chandrasekhar:1964zza,Chandrasekhar:1964zz}.

Before we proceed further, we observe that, in the Newtonian limit ($p\ll\rho$ and $\Phi, \Lambda\ll1$), it reduces to
\begin{equation}
\omega^2 \rho \xi=\frac{2(N-1)}{r}\frac{{\rm d}p}{{\rm d}r}\xi-\left[\frac{\gamma p}{r^{N-1}}\left(r^{N-1} \xi\right)'\right]',
\end{equation}
which is actually the pulsation equation (when $N=3$) in Ref.~\cite{Shapiro:1983du} by perturbing the Euler's equation in Newtonian Gravity (NG).
This implies the \emph{critical adiabatic index} (see App.~\ref{app:NGcritical} for derivation)
\begin{equation}
\boxed{
\gamma_{\rm cr}=2\left(1-\frac{1}{N}\right).
}
\end{equation}

To have a stable configuration, the perturbation cannot grow without bound, meaning the eigenfrequency $\omega$ must be real, in other words, $\omega^2>0$.
It follows from App.~\ref{app:NGcritical} that the pressure-averaged $\langle\gamma\rangle>\gamma_{\rm cr}$. For relativistic (non-relativistic) ideal fluids, this implies the spatial dimensions must be $N<3~(N<4)$ in order to have a stable sphere. From this viewpoint, the privilege of (3+1) dimensions is manifest because the fluid sphere \emph{is stable but not too stable.} However, in (2+1) dimensions, it is too stable because $\gamma_{\rm cr}=1<1.5~(2)=\gamma$ as always for an ultra-relativistic (non-relativistic) fluid. 
Nevertheless, the ``pressure effect'' is crucial in GR because the whole energy-momentum should be taken as a single entity, \emph{not only the energy density but also the pressure is sourcing gravity}.

To determine the critical adiabatic index in the full relativistic context, we perform the integration over $r$ with $\xi$ over the sphere with proper measure $r^{N-1}e^{\Phi+\Lambda}$, and integration by parts for the term with $\gamma$ in the integrand, resulting in
\begin{align}\label{seceq:puls2}
\nonumber
&\omega^2\int e^{3\Lambda-\Phi}(\rho+p)r^{N-1}{\xi}^{2}{\rm d}r=2(N-1)\int e^{\Phi+\Lambda}r^{N-2}\frac{{\rm d}p}{{\rm d}r}{\xi}^{2}{\rm d}r\\
\nonumber
&+\int e^{3\Phi+\Lambda}\frac{\gamma p}{r^{N-1}}\left[(r^{N-1}e^{-\Phi}\xi)'\right]^2{\rm d}r-\int e^{\Phi+\Lambda}\left(\frac{{\rm d}p}{{\rm d}r}\right)^2\frac{r^{N-1}\xi^{2}}{\rho+p}{\rm d}r\\
&+\frac{2\kappa_N}{N-1} \int e^{3\Lambda+\Phi}p(\rho+p)r^{N-1}\xi^{2}{\rm d}r.
\end{align}

By Rayleigh-Ritz principle (see App.~\ref{app:ortho}), $\omega^2\leq0$ signals the instability of the given configuration, and thus determines the critical adiabatic index $\gamma_{\rm cr}$ at $\omega^2=0$.  To see the relativistic corrections to NG, on the RHS of Eq.~(\ref{seceq:puls2}), we perform (i) integration by parts for the first term and with Eq.~(\ref{seceq:ttrr}) to replace $\Phi'+\Lambda'$; (ii) replacement of $p'$ by Eq.~(\ref{seceq:tov}) in the third term; and (iii) use of Eq.~(\ref{seceq:rr}) to replace $\Phi'$ after choosing the trial function $\xi(r)=re^\Phi$~\cite{Chandrasekhar:1964zz}. After arranging all the terms with care and setting RHS of Eq.~(\ref{seceq:puls2}) equal to zero, one obtains
\begin{align}\label{seceq:gammacrit}
\nonumber
\gamma_{\text{cr}}&\equiv2\left(1-\frac{1}{N}\right)+\frac{\int e^{3\Phi+\Lambda}[8(N-1)p+(N-2)(e^{2\Lambda}-1)(\rho+p)](N-2)(e^{2\Lambda}-1)r^{N-1}{\rm d}r}{4N^2\int e^{3\Phi+\Lambda}pr^{N-1}{\rm d}r}\\
+&\frac{\kappa_N\int e^{3(\Phi+\Lambda)}[4(N-1)p+(N-2)(e^{2\Lambda}+1)(\rho+p)]pr^{N+1}{\rm d}r}{N^2(N-1)\int e^{3\Phi+\Lambda}pr^{N-1}{\rm d}r}+\frac{\kappa_{N}^2\int e^{3\Phi+5\Lambda}(\rho+p)p^2r^{N+3}{\rm d}r}{N^2(N-1)^2\int e^{3\Phi+\Lambda}pr^{N-1}{\rm d}r},
\end{align}
and
\begin{equation}\label{seceq:gammaavg}
\langle\gamma\rangle\equiv\frac{\int e^{3\Phi+\Lambda}\gamma pr^{N-1}{\rm d}r}{\int e^{3\Phi+\Lambda}pr^{N-1}{\rm d}r}
\end{equation}
the ``effective'' (pressure-averaged) adiabatic index of the fluid sphere.

\subsection{\label{sec:cos} The effect of cosmological constant}
Furthermore, if the \emph{cosmological constant} $\lambda\propto\rho_\lambda=-p_\lambda$ is included in the previous derivation, it turns out to be the Schwarzschild-Tangherlini spacetime~\cite{Tangherlini:1963bw}, and the pulsation equation becomes
\begin{align}\label{seceq:pulscos}
\nonumber
\omega^{2}e^{2(\Lambda-\Phi)}(\rho+p){\xi}=
&\frac{2(N-1)}{r}\frac{{\rm d}p}{{\rm d}r}\xi-e^{-(2\Phi+\Lambda)}\left[e^{3\Phi+\Lambda}\frac{\gamma p}{r^{N-1}}\left(r^{N-1}e^{-\Phi}\xi\right)'\right]'\\
&+\frac{2\kappa_N}{N-1} e^{2\Lambda}(p+p_\lambda)(\rho+p)\xi-\frac{1}{\rho+p}\left(\frac{{\rm d}p}{{\rm d}r}\right)^2\xi.
\end{align}

As we will see, the extra term $p_\lambda$ from the cosmological constant is significant to the stability condition. In addition, we follow the same procedure to obtain
\begin{align}\label{seceq:gammacritcos}
\nonumber
&\gamma_{\rm cr}\equiv2\left(1-\frac{1}{N}\right)
+\frac{\int e^{3\Phi+\Lambda}[8(N-1)p+(N-2)(e^{2\Lambda}-1)(\rho+p)](N-2)(e^{2\Lambda}-1)r^{N-1}{\rm d}r}{4N^2\int e^{3\Phi+\Lambda}pr^{N-1}{\rm d}r}\\
\nonumber
&+\frac{\kappa_N\int e^{3(\Phi+\Lambda)}\left\{[4(N-1)p+(N-2)(e^{2\Lambda}+1)(\rho+p)](p+p_\lambda)-2(N-1)p_\lambda(\rho+p)\right\} r^{N+1}{\rm d}r}{N^2(N-1)\int e^{3\Phi+\Lambda}pr^{N-1}{\rm d}r}\\
&+\frac{\kappa_{N}^2\int e^{3\Phi+5\Lambda}(\rho+p)(p+p_\lambda)^2r^{N+3}{\rm d}r}{N^2(N-1)^2\int e^{3\Phi+\Lambda}pr^{N-1}{\rm d}r},
\end{align}
and $\langle\gamma\rangle$ is also given by Eq.~(\ref{seceq:gammaavg}).
The second term on RHS of Eq.~(\ref{seceq:gammacritcos}) tends to destabilize the fluid sphere due the ``pressure effect'' of the fluid; while the third and fourth terms depend on its \emph{competition} with cosmological constant $\lambda$. \emph{Qualitatively}, $\lambda<0$ tends to stabilize the sphere; and $\lambda>0$ does the opposite. In particular, we note that the impact of cosmological constant on the Chandrasekhar instability is opposite to the Antonov instability (gravothermal catastrophe)~\cite{Axenides:2012bf}.
The expression is fully relativistic in its own right, though it would be indicative to see the post-Newtonian expansion in a particular model. Therefore, we apply this result  to the homogeneous model in the next section.

\section{\label{sec:homogeneous}Homogeneous Fluid Solutions}
The total fluid mass of homogeneous density is
\begin{equation}
\mathcal{M}=\frac{\kappa_N}{N(N-1)}\rho R^N
\end{equation}
thus we can write
\begin{equation}
\kappa_N\rho=\frac{N(N-1)\mathcal{M}}{R^N}.
\end{equation}

On the other hand, the definition of cosmological constant is ambiguous up to some factor depending on the space dimensionality.
We adopt the convention based on $R_{\lambda\mu\nu\rho}=\lambda(g_{\lambda\nu}g_{\mu\rho}-g_{\lambda\rho}g_{\mu\nu})$ for a space of constant curvature to define the cosmological constant in ($N$+1) dimensions from the vacuum Einstein equations $G_{\mu\nu}=-\frac{N(N-1)}{2}\lambda g_{\mu\nu}$, thus
\begin{equation}
\kappa_N\rho_\lambda=\frac{N(N-1)}{2}\lambda=-\kappa_N p_\lambda,
\end{equation}
where $\lambda=\pm1/\ell^2$, with $\ell$ the curvature of radius, depends on the positiveness of the scalar curvature.
Then, the TOV equation Eq.~(\ref{seceq:tov}) can be solved analytically:
\begin{subequations}
\begin{equation}
\frac{p(r)}{\rho}
=\frac{ [(N-2)(\mathcal{M}/R^{N-2})-\lambda R^2]\left[ e^{-\Lambda(r)} - e^{-\Lambda(R)}\right] }{ N(\mathcal{M}/R^{N-2})~e^{-\Lambda(R)} - [(N-2)(\mathcal{M}/R^{N-2})-\lambda R^2]e^{-\Lambda(r)} }
\end{equation}
and
\begin{equation}
e^{\Phi(r)}
=\frac{ N(\mathcal{M}/R^{N-2})~e^{-\Lambda(R)} - [(N-2)(\mathcal{M}/R^{N-2})-\lambda R^2]e^{-\Lambda(r)} }{ 2\mathcal{M}/R^{N-2}+\lambda R^2 },
\end{equation}
where
\begin{equation}
e^{-\Lambda(r)}
=\sqrt{1-\left(\frac{2\mathcal{M}}{R^{N-2}}+\lambda R^2\right)\frac{r^2}{R^2}}.
\end{equation}

The solution is parameterized by the \emph{compactness parameter} $\mathcal{C}_N\equiv\mathcal{M}/R^{N-2}$, the \emph{curvature parameter} $\lambda R^2$ and the space dimensionality $N\geq2$. The parametrization makes sense only if $\rho\neq0$ or the compactness $\mathcal{M}/R^{N-2}\neq0$. In addition, $(N-2)(\mathcal{M}/R^{N-2})-\lambda R^2>0$ is required to have $p(r)>0$.
\end{subequations}
Even though the homogeneous model is not so realistic, it captures the essence of some underlying physics. For example, the Buchdahl stability bound can be shown simply by demanding $p(0)<\infty$ and $e^{\Phi(0)}>0$:
\begin{equation}
\frac{N-1}{N^2}\left(1-\sqrt{1-\frac{N^2}{(N-1)^2}\lambda R^2}\right)
<
\frac{\mathcal{M}}{R^{N-2}}
<
\frac{N-1}{N^2}\left(1+\sqrt{1-\frac{N^2}{(N-1)^2}\lambda R^2}\right).
\end{equation}

Clearly, for $N=3,~\lambda=0$, it reduces to the familiar Buchdahl bound $0<9\mathcal{M}/4<R$ in (3+1) dimensions without cosmological constant. We also note that the lower bound is larger than zero if $\lambda>0$.
For more realistic models, assuming the decreasing monotonicity of density is sufficient to prove the Buchdahl bound, see App.~\ref{app:Buch} for a rigorous proof.
Nonetheless, when it comes to real stellar equilibrium, the instability might already trigger well before the Buchdahl bound.

In the post-Newtonian expansion with background curvature, Eq.~(\ref{seceq:gammacritcos}) leads to
\begin{subequations}\label{seceq:gammacrithomo}
\begin{equation}
\gamma_{\rm cr}
=\frac{\lambda R^2}{(N-2)\mathcal{M}/R^{N-2}-\lambda R^2}
+\sum_{j, k=0,1,...}f_{jk}^{(N)}\left(\frac{\mathcal{M}}{R^{N-2}}\right)^j \left(\lambda R^2\right)^k,
\end{equation}
where the second terms on the RHS are post-Newtonian corrections with coefficients $f_{jk}^{(N)}$ depending on the density distribution and spatial dimensions $N$, except that $f_{00}^{(N)}=2\left(1-1/N\right)$, the exact Newtonian result; the first term is a \emph{stabilizer/destabilizer} characterizing the \emph{competition} between the compactness and the background curvature, and it can be expanded as
\begin{equation}
\frac{\lambda R^2}{(N-2)\mathcal{M}/R^{N-2}-\lambda R^2}
=
\begin{cases}
\sum_{n=1}^\infty\left(\frac{\lambda R^2}{(N-2)\mathcal{M}/R^{N-2}}\right)^n 
\quad{\rm if}~\abs{\lambda R^2}<(N-2)\mathcal{M}/R^{N-2} \\
-1-\sum_{n=1}^\infty\left(\frac{(N-2)\mathcal{M}/R^{N-2}}{\lambda R^2}\right)^n
 \quad{\rm if}~\abs{\lambda R^2}>(N-2)\mathcal{M}/R^{N-2}.
\end{cases}
\end{equation}
\end{subequations}

We note that this term is always $-1$ for $N=2$ if $\lambda\neq0$, and the two limits ($\lambda\rightarrow0 \Leftrightarrow N\rightarrow2$) do not commute 
\[
0=\lim_{N\rightarrow2}\lim_{\lambda\rightarrow0}
\frac{\lambda R^2}{(N-2)\mathcal{M}/R^{N-2}-\lambda R^2}
\neq
\lim_{\lambda\rightarrow0}\lim_{N\rightarrow2}
\frac{\lambda R^2}{(N-2)\mathcal{M}/R^{N-2}-\lambda R^2}=-1
\]
reflecting the fact that GR has no Newtonian limit in (2+1) dimensions. Therefore, \emph{Einsteinian stars are even stabler than Newtonian stars in (2+1) as the critical adiabatic index is reduced by one unit in GR compared to NG.}

\section{\label{sec:numerics}Numerical Results}
The effect of cosmological constant cannot be neglected when it comes to stability, as it could stabilize the fluid sphere in higher dimensions ($N>3$) with $\lambda<0$ or destabilize it in lower dimensions ($N<3$) with $\lambda>0$.
However, GR in lower dimensions, (1+1) has no dynamics (vacuum), and (2+1) has no Newtonian limit. In particular, $\lambda<0$ is required to have the Ba{\~n}ados-Teitelboim-Zanelli (BTZ) black hole solution~\cite{Banados:1992wn} and stellar equilibrium~\cite{Cruz:1994ar}, it would then be interesting to see if the instability condition can be triggered in (2+1) dimensions.
In this section, we \emph{numerically} solve the marginal stable configurations and determine the critical compactness of homogeneous spheres in ($N$+1) dimensions. We first discuss (3+1) and higher and then (2+1) dimensions, respectively.

\subsection{\label{sec:3+1}Fluid spheres in (3+1) and higher-dimensional spacetime}
\begin{figure}[htbp]%
\center
\includegraphics[height=1.0\textwidth]{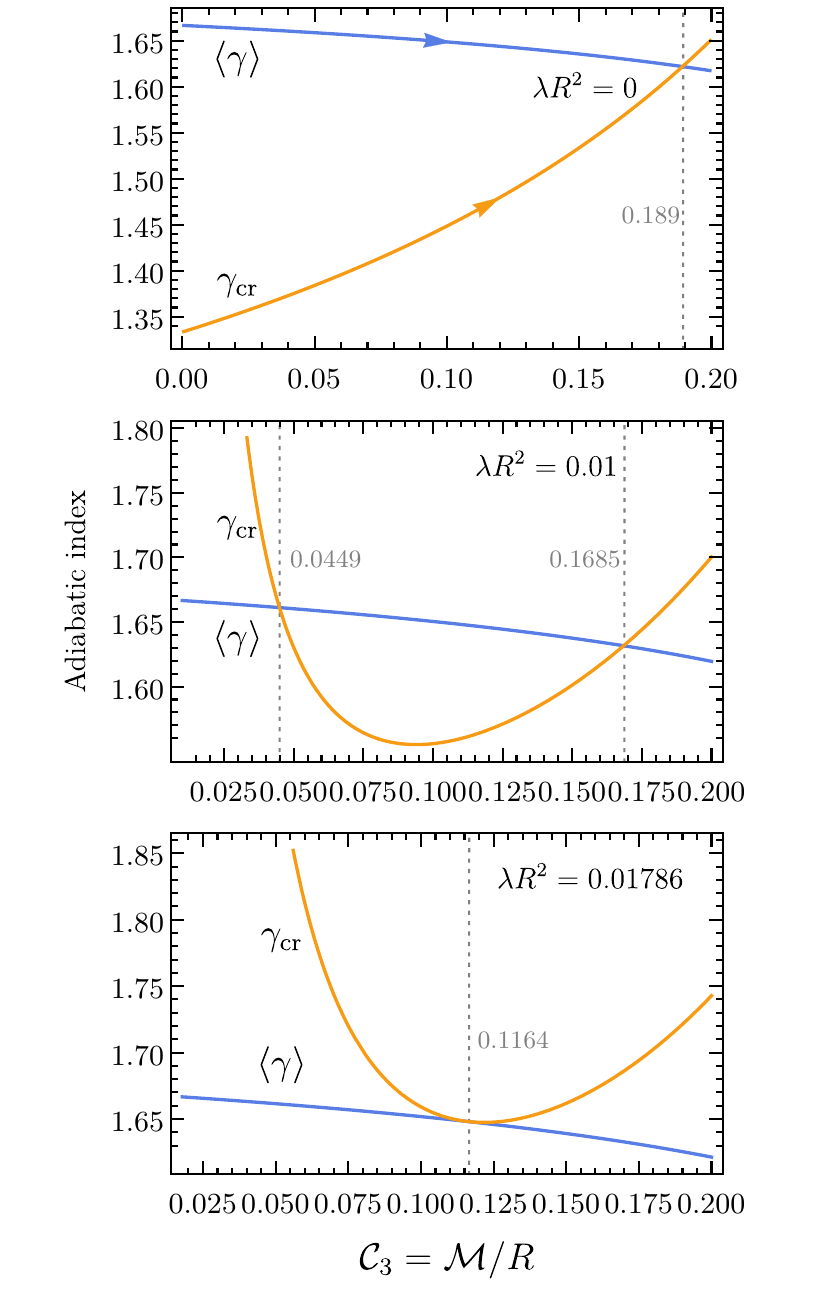}   
\caption{$N=3$: Pressure-averaged and critical adiabatic indices v.s. compactness $\mathcal{C}_3$ given $\lambda\geq0$. The configurations are unstable if $\langle\gamma\rangle<\gamma_{\rm cr}$, and the instabilities will set in at critical points $\langle\gamma\rangle=\gamma_{\rm cr}$. In the case of zero cosmological constant $\lambda R^2=0$ (\emph{top}), the arrows on the lines of adiabatic indices exhibit the directions when the fluid sphere is being compressed while keeping $\mathcal{M}$ fixed. There is only \emph{one} critical point for the instability to be triggered at $\mathcal{C}_3=0.189$.  On the other hand, if $\lambda>0$ is turned on (no matter how small it is), \emph{two} critical points will present. It is shown that, for $\lambda R^2=0.01$ (\emph{middle}), the stable region is bounded between $\mathcal{C}_3=0.0449$ and $0.1685$ and shrinks as $\lambda R^2$ increases until $\lambda R^2=0.01786$ (\emph{bottom}), at which $\mathcal{C}_3=0.1164$, the two critical points, become degenerate. In Tab.~\ref{tab:I}, we list the corresponding compactness $\mathcal{C}_3$ and central velocity dispersion $v_c$ of stable regions with various $\lambda R^2\geq0$. }
\label{fig:n3_adiabatic_plus}
\end{figure}

Without cosmological constant, a (3+1)-dimensional homogeneous ideal fluid becomes unstable as $\gamma_{\rm cr}=\langle\gamma\rangle=1.6219$, and the critical central velocity dispersion $v_c\equiv v(0)=0.681433$ with the critical compactness $\mathcal{C}_3=\mathcal{M}/R=0.189$, see Fig.~\ref{fig:n3_adiabatic_plus} (\emph{top}).
A positive cosmological constant tends to destabilize the sphere owing to the extra energy density and reduced pressure. Interestingly, in Fig.~\ref{fig:n3_adiabatic_plus} (\emph{middle}) there are \emph{two critical points} as $\lambda>0$ is turned on. In reality, if the configuration is not sufficiently compact, the fluid is unstable, and it tends to further contract until it transitions into a stable configuration~\cite{Feng:2022fuk}. However, the stable region shrinks as $\lambda R^2$ increases, and it could directly form a black hole if the stable region vanishes, see Tab.~\ref{tab:I}.

The stable configurations are bounded by the \emph{two critical points} up to $\lambda R^2=0.01786$, above which there is no stable configuration, see Fig.~\ref{fig:n3_adiabatic_plus} (\emph{bottom}). At this \emph{degenerate} critical point, $\mathcal{C}_3=\mathcal{M}/R=0.1164$, we can eliminate the dependence of $R$ to get $\mathcal{M}\sqrt{\lambda}=0.01556$, or $\mathcal{M}=0.01556\ell c^2/G_3$ with $G_3$ and $c$ restored. Now, given the cosmological constant observed~\cite{Planck:2015fie,Planck:2018vyg,Prat:2021xlz}, $\ell\sim10^{61}\ell_{\rm Pl}$, where $\ell_{\rm Pl}$ is the Planck length in (3+1) dimensions, there is no stable stellar equilibrium for $\mathcal{M}\gtrsim0.01556\ell c^2/G_3\sim10^{21}{\rm~M}_\odot$. Therefore, a virialized mass sphere should be much smaller $10^{21}{\rm~M}_\odot$ in order to have long-lived hydrostatic equilibrium before it can trigger the black hole formation.
Curiously, this number is two orders of magnitude larger to the maximal Jeans mass (the threshold that a gas cloud can clump into gravitationally bound states) just prior to the recombination of hydrogen $M_{\rm J}\simeq1.2\times10^{16}\left(\Omega_{b,0}h^2\right)^{-2}~{\rm M}_\odot
\simeq10^{19}~{\rm M}_\odot$ with the current baryon abundance $\Omega_{b,0}h^2\simeq0.0224$~\cite{Mo:2010ga,Planck:2018vyg}.
As the horizon mass of hydrogen is always less than the Jeans mass before recombination, structures can form only after recombination~\cite{Kolb:1990vq}.
Therefore the stable upper bound  $10^{21}~{\rm M}_\odot$ seems delicately protected in our universe.

\begin{figure}[htbp]%
\center
\includegraphics[height=0.42\textwidth]{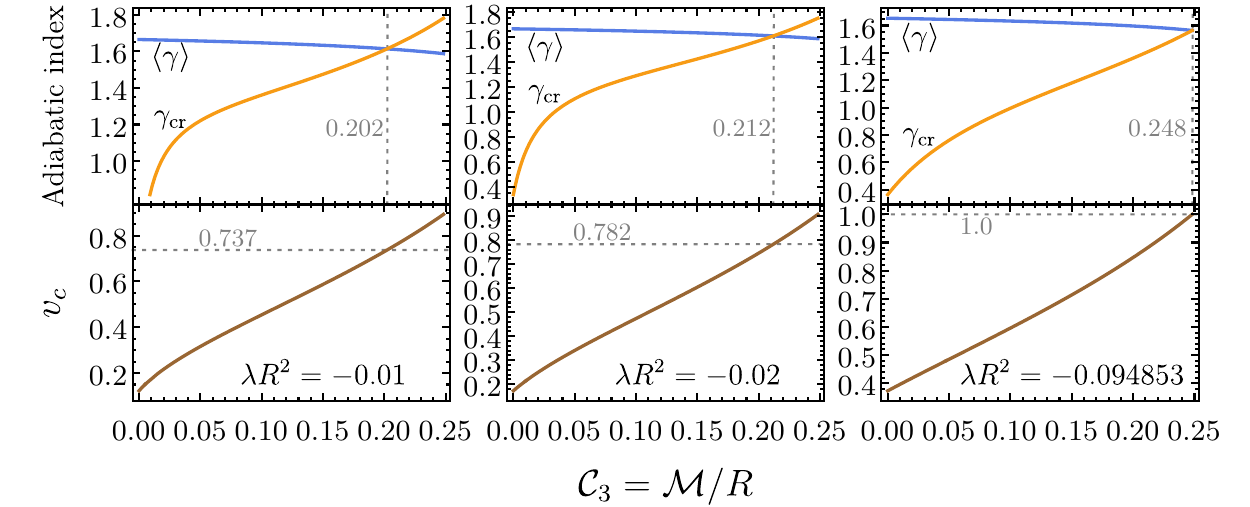}   
\caption{$N=3$: Pressure-averaged \& critical adiabatic indices (top panels), and central velocity dispersion $v_c$ (bottom panels) v.s. compactness $\mathcal{C}_3$ given $\lambda<0$. The configurations are unstable if $\langle\gamma\rangle<\gamma_{\rm cr}$, and the instabilities will set in at critical points $\langle\gamma\rangle=\gamma_{\rm cr}$. There is only one critical point for $\lambda<0$. For $\lambda R^2=-0.01$ (\emph{left}), the instability is to be triggered at $\mathcal{C}_3=0.202$ with $v_c=0.737$; while for $\lambda R^2=-0.02$ (\emph{middle}), it occurs at $\mathcal{C}_3=0.212$ with $v_c=0.782$. Compared to $\lambda R^2=0$ in Fig.~\ref{fig:n3_adiabatic_plus} (\emph{top}), it becomes harder to trigger the instabilities as higher $\mathcal{C}_3$ (thus $v_c$) is required if $\lambda$ is more negative until $\lambda R^2=-0.094853$ (\emph{right}), at which $\mathcal{C}_3=0.248$ with the causal limit $v_c=1$. Beyond this point, no physical configuration can trigger the instability on the grounds of causality. In Tab.~\ref{tab:II}, we list the critical points at causal limits for $N=3,4,5,6,$ and $7$. }
\label{fig:n3_adiabatic_minus}
\end{figure}

On the other hand, a negative cosmological constant tends to stabilize the sphere due to the extra pressure and reduced energy density. Compared to zero and positive cosmological constant, the critical compactness becomes larger in order to trigger the collapse. There is \emph{only one critical point} for a given compactness down to $\lambda R^2=-0.094853$, below which there is no physical solution due to causality ($v_c\leq1$).  At this critical point, $\mathcal{C}_3=0.248$, see Fig.~\ref{fig:n3_adiabatic_minus}.

For $N>3$, the fluid is genuinely unstable in NG. In GR, however, the fluid becomes stabilized if $\lambda<0$ is turned on. For $N=4,5,6$, and $7$, we also identify the lower bound of $\lambda R^2$ in Tab.~\ref{tab:II} that the fluid can still collapse into a black hole under the causal limit $v_c=1$.

\setlength{\tabcolsep}{15pt} 
\renewcommand{\arraystretch}{1} 
\begin{table}
  \caption{$N=3$: Stable regions $\langle\gamma\rangle\geq\gamma_{\rm cr}$ with various $\lambda R^2>0$. There is no stable configuration for $\lambda R^2\gtrsim0.01786$.}
 \resizebox{\columnwidth}{!}{%
 \begin{tabular}{l c c c}
  \hline\hline
  $\lambda R^2$  & $\mathcal{C}_3=\mathcal{M}/R$  &  $v_c=v(0)$ &  $\langle\gamma\rangle$ \\ [0.5ex] 
  \hline\hline
$0$ & $0\textup{--}0.189057$ & $0\textup{--}0.681433$ & $1.66667\textup{--}1.62190$ \\
\hline
$0.010$ & $0.044912\textup{--}0.16846$ & $0.240284\textup{--}0.604073$ & $1.66116\textup{--}1.63163$ \\
$0.015$ & $0.075802\textup{--}0.149753$ & $0.328969\textup{--}0.541498$ & $1.65634\textup{--}1.63857$ \\
$0.016$ & $0.084228\textup{--}0.14377$ & $0.352167\textup{--}0.522649$ & $1.65483\textup{--}1.64051$ \\
$0.017$ & $0.09501\textup{--}0.13543$ & $0.3818\textup{--}0.497134$ & $1.65275\textup{--}1.64302$ \\
$0.0175$ & $0.10278\textup{--}0.12889$ & $0.403279\textup{--}0.47766$ & $1.65113\textup{--}1.64485$ \\
$0.0176$ & $0.10485\textup{--}0.12705$ & $0.409033\textup{--}0.472255$ & $1.65069\textup{--}1.64534$ \\
$0.0177$ & $0.10738\textup{--}0.12475$ & $0.416092\textup{--}0.465544$ & $1.65013\textup{--}1.64594$ \\
$0.0178$ & $0.1109\textup{--}0.1215$ & $0.425964\textup{--}0.456132$ & $1.64933\textup{--}1.64678$ \\
$0.01785$ & $0.1142\textup{--}0.1183$ & $0.435279\textup{--}0.446946$ & $1.64856\textup{--}1.64757$ \\
\hline
$0.01786$ & $0.1164$ & $0.441523$ & $1.64804$ \\
[0.5ex]
  \hline\hline
 \end{tabular}
 }
 \label{tab:I}
\end{table}

\setlength{\tabcolsep}{35pt} 
\renewcommand{\arraystretch}{1} 
\begin{table}
  \caption{Critical points for $N=3,4,5,6,$ and $7$ with $\lambda<0$ at causal limit $v_c=1$.}
 \resizebox{\columnwidth}{!}{%
 \begin{tabular}{c c c c}
  \hline\hline
  $N$  &  $\mathcal{C}_N$  & $\lambda R^2$ & $\langle\gamma\rangle=\gamma_{{\rm cr}}$ \\ [0.5ex] 
  \hline\hline
$3$ & $0.248179$ & $-0.094853$ & $1.56387$ \\
$4$ & $0.117505$ & $-0.134605$ & $1.43352$ \\
$5$ & $0.062846$ & $-0.151149$ & $1.35328$ \\
$6$ & $0.037099$ & $-0.154395$ & $1.29861$ \\
$7$ & $0.023595$ & $-0.151406$ & $1.25884$ \\
[0.5ex]
  \hline\hline
 \end{tabular}
 }
 \label{tab:II}
\end{table}

\subsection{\label{sec:2+1}Fluid disks in (2+1)-dimensional spacetime}
It is well known that (2+1) GR has no local degrees of freedom (locally flat), thus no gravitational wave (or graviton) can propagate. This means that particles do not gravitate if they are static in the (2+1)-dimensional spacetime~\cite{Giddings:1983es,Deser:1983tn,Deser:1983nh}.
On the other hand, the collective behavior of particles demands the fluid description under the influence of gravity~\cite{Giddings:1983es}. To have hydrostatic equilibrium, a negative cosmological constant $\lambda=-1/\ell^2$ is to guarantee not only hydrostatic equilibrium~\cite{Cruz:1994ar} (the pressure is monotonically decreasing) but also permit a black hole solution (BTZ) in (2+1) dimensions. 
The metric interior to the fluid disk turns out to be
\begin{equation}
{\rm d}s_{\rm disk}^2=-e^{2\Phi(r)}{\rm d}t^2+e^{2\Lambda(r)}{\rm d}r^2+r^2{\rm d}\phi^2
\quad{\rm with}\quad
e^{-2\Lambda(r)}=1-2M(r)+\frac{r^2}{\ell^2}.
\end{equation}
Note that, in (2 + 1) dimensions, the integration constant in $e^{-2\Lambda(r)}$ is arbitrary, but it can always be normalized to ``unity'' by adjusting the natural mass scale $G_2$.
On the other hand, the non-rotating BTZ metric reads
\begin{equation}
{\rm d}s_{\rm BTZ}^2=-f(r){\rm d}t^2+\frac{{\rm d}r^2}{f(r)}+r^2{\rm d}\phi^2
\quad{\rm with}\quad
f(r)=-\mathcal{M}_\text{BTZ}+\frac{r^2}{\ell^2}.
\end{equation}
To match spacetime of the fluid interior to the BTZ exterior, the junction condition at the fluid radius $R$ leads to the relation of the BTZ mass and the fluid mass 
\begin{equation}
\boxed{
\mathcal{M}_{\rm BTZ}=2\mathcal{M}-1.
}
\end{equation}

\begin{figure}%
\centering
\subfigure[~$0<\mathcal{M}\leq0.5$ ($-1<\mathcal{M}_{\rm BTZ}\leq0$)]{%
\label{fig:btzminus}%
\includegraphics[height=2.15in]{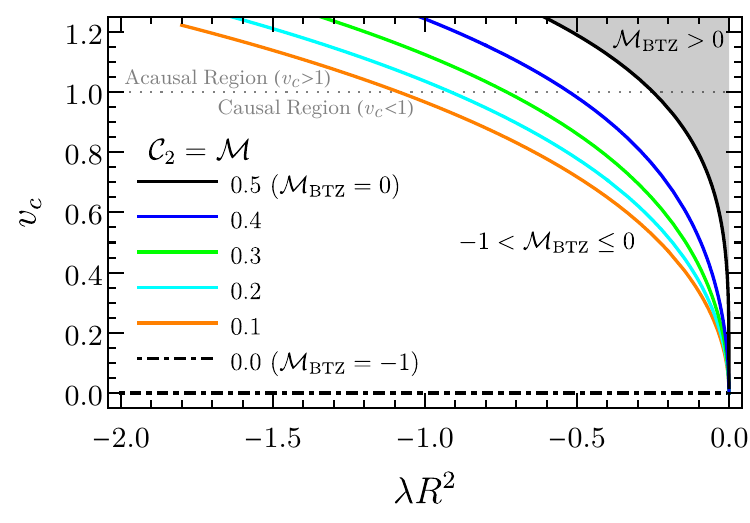}}%
\qquad
\subfigure[~$\mathcal{M}>0.5$ ($\mathcal{M}_{\rm BTZ}>0$)]{%
\label{fig:btzplus}%
\includegraphics[height=2.15in]{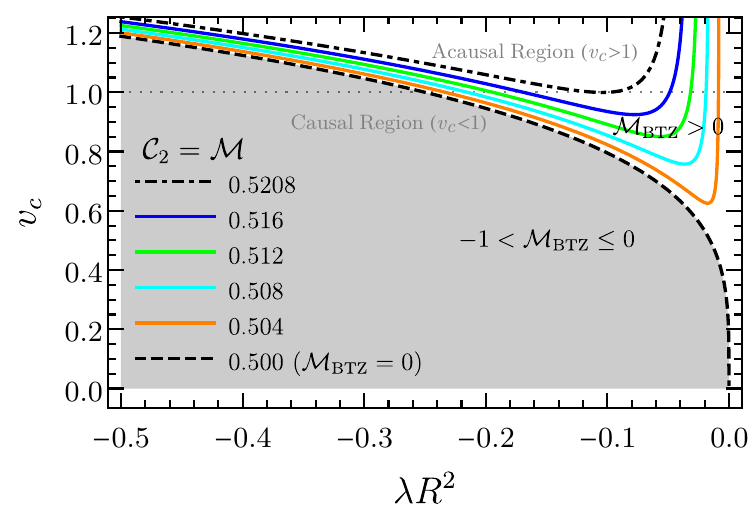}}%
\caption{$v_c-\lambda R^2$ phase diagrams of homogeneous fluid disks in (2+1) dimensions with the total fluid mass (a) $0<\mathcal{M}\leq0.5$; and (b) $\mathcal{M}>0.5$. The BTZ bound state (AdS spacetime) $\mathcal{M}_{\rm BTZ}=-1$ ($\mathcal{M}=0$ only if $R=0$ as $\rho\neq0$) is separated from the mass spectrum. The fluid solution for $\mathcal{M}\gtrsim0.5208$ is forbidden regarding causality. In App.~\ref{app:n2tables}, we list the corresponding $\lambda R^2$ at causal limit $v_c=1$ for various $0<\mathcal{M}\leq0.5$ in Tab.~\ref{tab:III} and $\mathcal{M}>0.5$ in Tab.~\ref{tab:IV}, respectively. The minima of $v_c$ for $\mathcal{M}>0.5$ are shown in Tab.~\ref{tab:V}. } 
\end{figure}

We note that the ADM mass in (2+1) is the BTZ mass~\cite{Banados:1992gq,Brown:1986nw,Brown:1994gs}, rather than the fluid mass. The phase diagrams of homogeneous fluid configurations with $0<\mathcal{M}\leq0.5$ and $\mathcal{M}>0.5$ are shown in Figs.~\ref{fig:btzminus} and~\ref{fig:btzplus}, respectively.
The absence of fluid $\mathcal{M}=0$ corresponds to the BTZ bound state (anti-de Sitter space) $\mathcal{M}_{\rm BTZ}=-1$, which is separated from the mass spectrum.
The fluid mass $\mathcal{M}=0.5$ is the BTZ ground (vacuum) state $\mathcal{M}_{\rm BTZ}=0$. For $0<\mathcal{M}<0.5$, they are a sequence of states of naked conical singularity~\cite{Banados:1992gq}, $-1<\mathcal{M}_{\rm BTZ}<0$, if the fluid were to collapse. Further, the threshold to have BTZ excited state $\mathcal{M}_{\rm BTZ}>0$ is $\mathcal{M}>0.5$.  It is interesting to note that the central velocity dispersion $v_c$ is always monotonically increasing in $\abs{\lambda R^2}$ for $0<\mathcal{M}\leq0.5$, while there are minima of $v_c\neq0$ for $\mathcal{M}>0.5$, and no causal solution if $\mathcal{M}\gtrsim0.5208$. This manifests that the two phase diagrams are separated by the \emph{phase boundary} $\mathcal{M}=0.5$ (or $\mathcal{M}_{\rm BTZ}=0$).
 
By Chandrasekhar's criterion at the critical point $\langle\gamma\rangle=\gamma_{\rm cr}$, we can examine if the BTZ excited states and the BTZ bound states (naked singularities) can result from a collapsing fluid. 
For a homogeneous disk, the critical adiabatic index, Eq.~(\ref{seceq:gammacrithomo}), reduces to 
\begin{align}\label{seceq:gammacrithomo_n2}
\gamma_{\rm cr}
=-1&+\sum_{j, k=0,1,...}f_{jk}^{(2)}\mathcal{M}^j \left(\lambda R^2\right)^k
=\left(-\frac{1}{3}\left(\lambda R^2\right)-\frac{11}{72}\left(\lambda R^2\right)^2-\frac{37}{432}\left(\lambda R^2\right)^3+...\right)
\nonumber\\
&+\left(-\frac{3}{4}\left(\lambda R^2\right)-\frac{3}{4}\left(\lambda R^2\right)^2-\frac{125}{192}\left(\lambda R^2\right)^3+...\right)\mathcal{M}
\nonumber\\
&+\left(-\frac{19}{12}\left(\lambda R^2\right)-\frac{733}{288}\left(\lambda R^2\right)^2-\frac{439}{144}\left(\lambda R^2\right)^3+...\right)\mathcal{M}^2
\nonumber\\
&+\left(-\frac{157}{48}\left(\lambda R^2\right)-\frac{133}{18}\left(\lambda R^2\right)^2-\frac{821}{72}\left(\lambda R^2\right)^3+...\right)\mathcal{M}^3
+\mathcal{O}\left(\mathcal{M}^4\right).
\end{align}

\begin{figure}[htbp]%
\center
\includegraphics[height=0.4\textwidth]{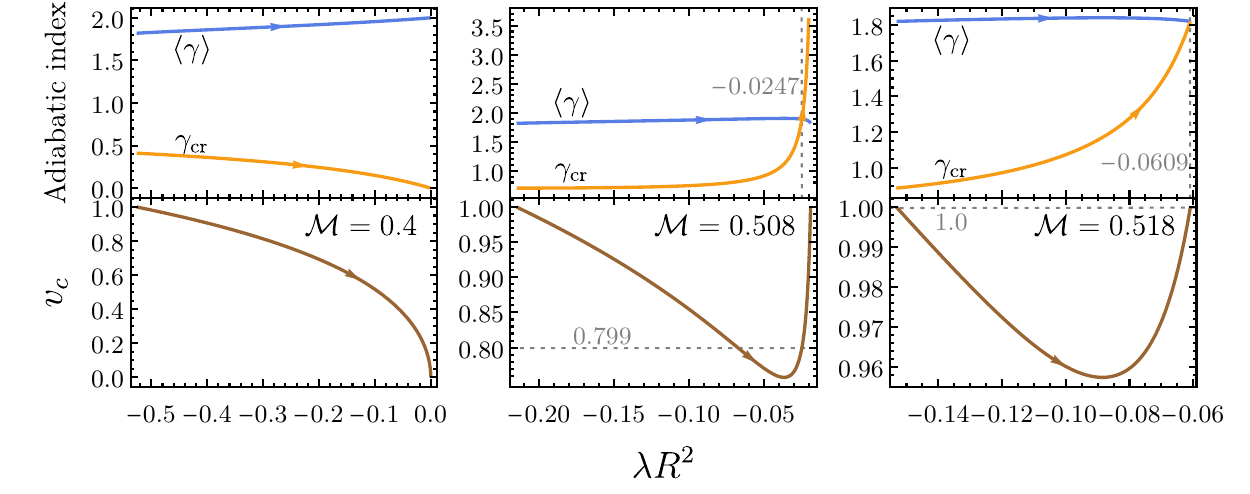}   
\caption{$N=2$: Pressure-averaged and critical adiabatic indices (top panels), and central velocity dispersion $v_c$ (bottom panels) v.s. curvature parameter $\lambda R^2$. The configurations are unstable if $\langle\gamma\rangle<\gamma_{\rm cr}$, and the instabilities will set in at critical points $\langle\gamma\rangle=\gamma_{\rm cr}$. The arrows on the lines of adiabatic indices exhibit the directions when the fluid sphere is being compressed while keeping $\mathcal{C}_2=\mathcal{M}$ and $\lambda$  fixed. For $\mathcal{M}\leq0.5$, there is no crossing of the two indices, thus no instability will be triggered. For example, for $\mathcal{M}=0.4$ (\emph{left}), $\gamma_{\rm cr}$ ($\langle\gamma\rangle$) is maximal (minimal) at $v_c=1$ and decreases down (increases up) to zero (two) as $R$ decreases and the two indices never cross. On the contrary, instabilities can occur for $\mathcal{M}>0.5$. For $\mathcal{M}=0.508$ (\emph{middle}) the instability can set in at $\lambda R^2=-0.0247$ with $v_c=0.799$ ; however, for $\mathcal{M}=0.518$ (\emph{right}), it is at $\lambda R^2=-0.0609$ with $v_c=1$; no instability can be triggered if $\mathcal{M}\gtrsim0.518$ regarding causality.}
\label{fig:n2_adiabatic}
\end{figure}

It ostensibly starts from ``zero'' rather than ``one'' in NG as $\lambda R^2\rightarrow0$.  However, Eq.~(\ref{seceq:gammacrithomo_n2}) is \emph{not necessarily} convergent to zero as $\lambda R^2\rightarrow0$, and the convergence really relies on the fluid mass $\mathcal{M}$. In fact, $\gamma_{\rm cr}\rightarrow0$ as $\lambda R^2\rightarrow0$ \emph{only if} $\mathcal{M}\leq0.5$ when the fluid is being compressed. Thus, the presence of the negative cosmological constant makes the relativistic instability hardly be triggered for $\mathcal{M}\leq0.5$. In Fig.~\ref{fig:n2_adiabatic} (\emph{left}) we take $\mathcal{M}=0.4$, for example, and it is always $\gamma_{\rm cr}<\langle\gamma\rangle$, thus stable, under causal region $v_c\leq1$.  On the other hand, the instability could set in if the fluid disk exceeds, \emph{no matter how tiny amount}, the threshold $\mathcal{M}=0.5$ (or $\mathcal{M}_{\rm BTZ}=0$). Nevertheless, some external agent must compress the fluid to make it unstable if $\lambda$ is fixed. In Fig.~\ref{fig:n2_adiabatic} (\emph{middle}), for $\mathcal{M}=0.508$ the instability can set in at $\lambda R^2=-0.0247$ with central velocity dispersion $v_c=0.799$. For higher $\mathcal{M}$ both $\abs{\lambda R^2}$ and $v_c$ increase at the critical point of instability until $\mathcal{M}=0.518$ at $\lambda R^2=-0.0609$ with $v_c=1$. Beyond this mass, there is no unstable configuration under causal range $v_c\leq1$. In App.~\ref{app:n2tables}, we also list the critical points ($\lambda R^2, v_c$) of instability for various masses $0.5<\mathcal{M}\lesssim0.518$ in Tab.~\ref{tab:VI}.

\section{\label{sec:discuss}discussions and implications}
We have derived Chandrasekhar's criterion in ($N$+1) spacetime with and without cosmological constant. As an illustration, we take the homogeneous solution to determine the instability provided that the sphere is composed of an ideal monatomic fluid.

In (3+1) spacetime, the privileged position is manifest as it is the {marginal dimensionality} in which the fluid sphere is stable but not too stable to trigger the onset of gravitational collapse. In particular, it is the \emph{unique} dimensionality that allows stable hydrostatic equilibrium with a positive cosmological constant. For higher spatial dimensions, the fluid sphere is genuinely unstable either in the context of NG or GR. However, a negative cosmological constant can stabilize it.

In (2+1) spacetime the effect of negative cosmological constant wins the relativistic effect so that it is too stable for a fluid disk of mass $0<\mathcal{M}\leq0.5$ to collapse into a naked singularity. This somewhat supports the Cosmic Censorship Conjecture~\cite{Penrose:1969pc}. However, the BTZ hole emergence is possible from a collapsing fluid under proper conditions. This is reasonable because, if it were not the case, the anti-de Sitter space $\mathcal{M}_{\rm BTZ}=-1$ (bound state) could be deformed \emph{continuously} into a BTZ black hole of mass $\mathcal{M}_{\rm BTZ}\geq0$ (vacuum or excited state) by growing the mass of the fluid disk from $\mathcal{M}=0$ to $\mathcal{M}\geq0.5$.

We now summarize the assumptions made implicitly in the results and the implications:
\\
\paragraph*{Assumptions:}
\begin{itemize}
\item First law of thermodynamics and equipartition theorem hold. 
\item Mass-energy dispersion relation $E=\sqrt{\mathbf{p}^2+m^2}$ is valid.
\item Einstein equations hold in ($N$+1) dimensions. 
\end{itemize}

The particles in a fluid sphere (or disk) are well thermalized at each local point in the spacetime 
such that the equipartition theorem holds; thus, the pressure is isotropic macroscopically. Microscopically, every particle in the fluid follows the mass-energy relation if the Lorentz symmetry is locally preserved. This is based on the assumption that particle states can be described by vectors in some irreducible representation of the Poincar\'e group; however, Lorentz symmetry is not necessary if the vectors describing particles can be generated from a vacuum vector with the help of local field operators~\cite{Borchers:1984yv}. In the fluid description, its adiabatic index can vary from $1+2/N$ to $1+1/N$ as particles go from non-relativistic to ultra-relativistic regimes as more and more gravitational energy is converted into the fluid.
On the gravity side, we restrict to only ``one time dimension'' since the hyperbolicity of field equations (in this note GR or its Newtonian limit) determines the causal structure~\cite{Tegmark:1997jg}, hence  the ``predictivity'' allows physicists to do theoretical physics. Moreover, in view of the Ostrogradsky theorem~\cite{Woodard:2015zca}, the field equations are at most second derivative because equations of motion with higher-order derivatives are in principle unstable or non-local. Both GR and NG are free of Ostrogradsky instability.

Although gravitational attraction of a fluid is getting weaker as space dimension $N$ increases, its pressure is even weaker when being compressed because the \emph{randomly} moving particles inside the fluid will have \emph{more directions} to go, and thus be easier to collapse. This is the reason why an ideal monatomic fluid is genuinely unstable in higher dimensions.
\\
\paragraph*{Implications:}
\begin{itemize}
\item Baby universe emerges from collapsing matter in a black hole. 
\item Spacetime dimensionality reshuffles in the reign of quantum gravity. 
\end{itemize}

We can reexamine the idea that the observable universe is the interior of a black hole~\cite{Pathria:1972,Good:1972,Easson:2001qf,Gaztanaga:2021qcw,Gaztanaga:2022gbd,Roupas:2022gee} existing as one of possibly many inside a larger parent universe, or multiverse. Since singularity is generic~\cite{Penrose:1964wq,Hawking:1965mf,Geroch:1966ur,Ellis:1968vy,Garfinkle:2003bb} in GR, some limiting curvature must exist~\cite{Frolov:1988vj,Frolov:1989pf} to avoid the singularity formation and transition to a baby universe. Several mechanisms, including extended GR with torsion~\cite{Poplawski:2010kb}, invoking stringy Hagedorn matter~\cite{Dubovsky:2012wk,Feng:2018jrh} to produce a bouncing solution~\cite{Biswas:2005qr,Biswas:2006bs}, braneworld scenario~\cite{Dvali:2000hr,Pourhasan:2013mqa}, or transition through an S-brane~\cite{Brandenberger:2021ken}, have been introduced to realize the idea. 

Suppose the space dimension reshuffling is a random process as collapsing matter into a black hole in a universe of arbitrary space dimension $N$. When the collapsing matter is squeezed down to the Planck scale (near the classical black hole singularity), it would emerge to a new universe with different space dimensions. 
The change of spacetime dimensions through phase transition near Planck scale might require some unknown theory of quantum gravity or string theory~\cite{Brandenberger:1988aj,Greene:2012sa}. Some inflation models that predict parts of exponentially large size having different dimensionalities~\cite{Linde:1988yp} might provide an alternative mechanism.
This process will repeat again and again until the new-born universe is just (3+1)-dimensional, which is stable but not too stable for the pristine gas in it to form complicated structures, including both black holes and stars. In any higher dimensions $N>3$, it is to too easy for matter to collapse into black holes; and in (2+1), there is no way to form black holes from self-gravitating fluid without external agents---\emph{Both are too barren to have complex structures.}

The above discussion reverberates the \emph{anthropic principle}~\cite{Barrow:1983,Barrow:1988yia,Smolin:1999,Susskind:2003kw,Linde:2002gj} in one way or another, though it only relies on the \emph{assumptions} regardless of the anthropic reasoning.
However, it tells \emph{more} than that because gravitationally bound states of monatomic fluid are genuinely unstable and bound to black holes immediately in space dimensions higher than three. Although a negative cosmological constant could stabilize the fluid spheres, it decelerates the expansion of the whole universe on a large scale. This leads to another issue on the anthropic bound of cosmological constant~\cite{Barrow:1988yia,Linde:2002gj,Bousso:2004}, even though the falsifiability has been criticized~\cite{Ball:2004,Smolin:2004yv}.  In (3+1), the anthropic bound on the positive cosmological constant~\cite{Weinberg:1987dv} argues that it could not be very large, or the universe would expand too fast for galaxies, stars (or us) to form. While once these local structures are detached from the background expansion, its \emph{smallness} also prevents the gravitationally bound state (typically of mass much less than $10^{21}~{\rm M}_\odot$) from forming a black hole immediately given the current cosmological constant observed.

Why $N=3$? Perhaps the more sensible question is not what makes (3+1) the preferred dimensionality~\cite{Momen:2011jc,Gonzalez-Ayala:2015xda,Brandenberger:1988aj,Greene:2012sa}, but, rather, \emph{why the physical principles (thermodynamics, Lorentz symmetry, GR,...) allow fecund universes to exist only in (3+1), which is the unique dimensionality that permits stable hydrostatic equilibrium with a positive cosmological constant.}

\section*{Acknowledgements}
The author acknowledges the Institute of Physics, Academia Sinica, for the hospitality during the completion of this work, and many colleagues there for discussions. The author is also grateful for helpful correspondence with Steve Carlip and Stanley Deser. This work is supported in part by the U.S. Department of Energy under grant No. DE-SC0008541.

\appendix
\section{\label{app:thermo}Thermodynamic identity}
Assuming the EoS $n=n(\rho, p)$, we have
\begin{equation}\label{appeq:thermo1}
\left(\frac{\partial{n}}{\partial\rho}\right)_{s}=\left(\frac{\partial{n}}{\partial\rho}\right)_{p}+\left(\frac{\partial{n}}{\partial{p}}\right)_{\rho}\left(\frac{\partial{p}}{\partial\rho}\right)_{s}
\end{equation}
and the first law of thermodynamics under adiabatic process leads to
\begin{equation}\label{appeq:thermo2}
\left(\frac{\partial{n}}{\partial\rho}\right)_{s}=\frac{n_{0}}{\rho_{0}+p_{0}}.
\end{equation}
Combining the above results, we obtain
\begin{equation}\label{appeq:thermo3}
\left(\frac{\partial{n}}{\partial\rho}\right)_{p}=\frac{n_{0}}{\rho_{0}+p_{0}}-\left(\frac{\partial{n}}{\partial{p}}\right)_{\rho}\left(\frac{\partial{p}}{\partial\rho}\right)_{s}
\end{equation}
where the subscript $s$ denotes an isentropic (adiabatic) process.

\section{\label{app:idealgas}Ideal monatomic fluid in $N$-dimensional space}
If we assume a fluid element of volume $\mathcal{V}$, and composed of $\mathcal{N}$ structureless point particles with the same mass $m$ in $N$-dimensional space, then
\begin{equation}
n(\mathbf{x})=\sum_{i=1}^{\mathcal{N}}\delta^{(N)}(\mathbf{x}-\mathbf{x}_{i})=\frac{\mathcal{N}}{\mathcal{V}},
\end{equation}
\begin{equation}
\rho(\mathbf{x})=\sum_{i=1}^{\mathcal{N}}E_{i}\delta^{(N)}(\mathbf{x}-\mathbf{x}_{i}),
\end{equation}
\begin{equation}
p(\mathbf{x})=\frac{1}{N}\sum_{i=1}^{\mathcal{N}}\frac{\mathbf{p}_{i}^2}{E_{i}}\delta^{(N)}(\mathbf{x}-\mathbf{x}_{i}),
\end{equation}
with the velocity of the $i-$th particle being $\mathbf{v}_{i}=\mathbf{p}_{i}/E_{i}$.
We have 
\begin{equation}
\mathbf{p}_{i}=\frac{m\mathbf{v}_{i}}{\sqrt{1-\mathbf{v}_{i}^2}},
\end{equation}
\begin{equation}
E_{i}=\frac{\abs{\mathbf{p}_{i}}}{\abs{\mathbf{v}_{i}}}=\frac{m}{\sqrt{1-\mathbf{v}_{i}^2}}.
\end{equation}

Given a distribution of these particles, we may take the root-mean-square speed as velocity dispersion by 
\begin{equation}
v\equiv v_{\text{rms}}=\sqrt{\frac{\mathcal{V}}{\mathcal{N}}\sum_{i}^{\mathcal{N}}\mathbf{v}_{i}^2\delta^{(N)}(\mathbf{x}-\mathbf{x}_{i})}.
\end{equation}

First, we express (root-mean-square speed averaged) energy density $\rho=\rho(p)$, by
\begin{equation}
p=\frac{1}{N}\sum_{i}^{\mathcal{N}}\frac{\mathbf{p}_{i}^2}{E_{i}}\delta^{(N)}(\mathbf{x}-\mathbf{x}_{i})=\frac{1}{N}\sum_{i}^{\mathcal{N}}E_{i}\mathbf{v}_{i}^2\delta^{(N)}(\mathbf{x}-\mathbf{x}_{i})=\frac{1}{N}\rho v^{2},
\quad{\rm thus}\quad
\rho=\frac{Np}{v^2}
\end{equation}

Second, we express $mn=mn(p)$ by
\begin{equation}
p=\frac{1}{N}\sum_{i}^{\mathcal{N}}\bigg(\frac{m\mathbf{v}_{i}}{\sqrt{1-\mathbf{v}_{i}^2}}\bigg)^2\frac{\sqrt{1-\mathbf{v}_{i}^2}}{m}\delta^{(N)}(\mathbf{x}-\mathbf{x}_{i})
=\frac{1}{N}\sum_{i}^{\mathcal{N}}\frac{\mathbf{v}_{i}^2}{\sqrt{1-\mathbf{v}_{i}^2}}m\delta^{(N)}(\mathbf{x}-\mathbf{x}_{i})=\frac{1}{N}\frac{v^{2}}{\sqrt{1-\tilde{v}^2}}mn,
\end{equation}
whence we can identify
\begin{equation}
\frac{1}{\sqrt{1-\tilde{v}^2}}mn=\rho,
\quad{\rm thus}\quad
mn=\frac{N\sqrt{1-\tilde{v}^2}}{v^2}p.
\end{equation}

The internal energy density
\begin{equation}
u\equiv\rho-mn=\frac{N(1-\sqrt{1-\tilde{v}^2})}{v^2}p\equiv(\gamma-1)^{-1}p,
\end{equation}
as $\tilde{v}\simeq v$, we find
\begin{equation}
\gamma(v)=1+\frac{v^{2}/N}{1-\sqrt{1-v^{2}}}=1+\frac{1+\sqrt{1-v^2}}{N}.
\end{equation}

As expected, in nonrelativistic limit ($v\rightarrow 0$), $\gamma\rightarrow 1+2/N$; in extremely relativistic limit ($v\rightarrow 1$), $\gamma\rightarrow 1+1/N$.

\section{\label{app:NGcritical}The critical adiabatic index in Newtonian Gravity}
Starting from the pulsation equation in NG
\begin{equation}
\omega^2 \rho \xi=\frac{2(N-1)}{r}\frac{{\rm d}p}{{\rm d}r}\xi-\left[\frac{\gamma p}{r^{N-1}}\left(r^{N-1} \xi\right)'\right]',
\end{equation}
we perform the integration over $r$ with $\xi$ over the sphere with proper measure $r^{N-1}$ (see App.~\ref{app:ortho} in the Newtonian limit $p\ll\rho$ and $\Phi, \Lambda\ll1$), and after integration by parts one obtains
\[
\omega^2\int\rho \xi^2 r^{N-1} {\rm d}r=\int\frac{\gamma p}{r^{N-1}}\left(r^{N-1} \xi\right)'^2 {\rm d}r-2(N-1)\int p\left(r^{N-2}\xi^2\right)' {\rm d}r.
\]

Once the trial function $\xi=r$ is chosen, we get
\[
\omega^2\int\rho r^{N+1} {\rm d}r=\int\left[N^2\gamma-2N(N-1)\right]pr^{N-1} {\rm d}r.
\]

By Rayleigh-Ritz principle, this implies the critical adiabatic index
\begin{equation}
\gamma_{\text{cr}}=2\left(1-\frac{1}{N}\right)
\end{equation}
with the pressure-averaged adiabatic index
\begin{equation}
\langle\gamma\rangle=\frac{\int\gamma p r^{N-1} {\rm d}r}{\int pr^{N-1} {\rm d}r}.
\end{equation}

As a result, $\langle\gamma\rangle>2(1-1/N)$ corresponds to stable oscillations ($\omega^2>0$); $\langle\gamma\rangle<2(1-1/N)$ corresponds to unstable collapse or explosion ($\omega^2<0$); and $\langle\gamma\rangle=2(1-1/N)$ is marginal stable ($\omega^2=0$). We note that this result is genuine and can be derived also by mode expansion of the trial function $\xi$ in~\cite{Arbanil:2019mae}.

\section{\label{app:ortho}The orthogonality relation \& Rayleigh-Ritz principle}
The orthogonality relation:
\begin{equation}
\int e^{2(\Lambda-\Phi)}(\rho+p)\xi^{(i)}\xi^{(j)}\sqrt{-g}{\rm d}r
=\int e^{3\Lambda-\Phi}(\rho+p)r^{N-1}\xi^{(i)}\xi^{(j)}{\rm d}r=0\quad (i\neq j),
\end{equation}
where $\xi^{(i)}$ and $\xi^{(j)}$ are the proper (eigen)solutions belonging to different characteristic values of $\omega^2$.
For Eq.~(\ref{seceq:puls1}) can be written as 
\begin{equation}
-\frac{\partial^{2}\xi}{\partial{t}^2}=\omega^2\xi=A\xi, 
\end{equation}
where $A$ is a linear differential operator which is self-adjoint, \emph{i.e.},
\begin{equation}
\langle\xi^{(i)}, A\xi^{(j)}\rangle=\langle A\xi^{(i)}, \xi^{(j)}\rangle,
\end{equation}
with the inner product $\langle\cdot, \cdot\rangle$ taken with a weight factor, $e^{3\Lambda-\Phi}(\rho+p)r^{N-1}$ in our case.  By \emph{Rayleigh-Ritz principle}, we have
\begin{equation}
\omega^2\leq\frac{\langle\xi, A\xi\rangle}{\langle\xi,\xi\rangle},
\end{equation}
for some chosen ``trial function'' $\xi$ (which need not be an eigensolution). Thus, if $\langle\xi, A\xi\rangle=0$, we obtain $\omega^2\leq 0$;
then $\xi$ grows without bound, and the perturbation is unstable.
Therefore, a ``sufficient'' condition for the onset of dynamical instability is that the RHS of Eq.~(\ref{seceq:puls2}) vanishes for the chosen $\xi$, which satisfies the required boundary conditions.

\section{\label{app:Buch}A rigorous proof on the Buchdahl bound}
The basic assumptions in the Buchdahl stability bound are:
\begin{itemize}
 \item The energy density $\rho$ is finite and monotonically non-increasing, \emph{i.e.}, ${\rm d}\rho/{\rm d}r\leq0$. 
 \item $e^{2\Phi}$ and $e^{2\Lambda}$ are positive definite, thus no horizon is present inside the fluid sphere.
 \end{itemize}
Including the cosmological constant $\lambda$, $\rho_\lambda$ and $p_\lambda$ are given by
$\kappa_N\rho_\lambda=\frac{N(N-1)}{2}\lambda=-\kappa_N p_\lambda$,
and the TOV equation Eq.~(\ref{seceq:tov}) along with the conservation law $p'=-(\rho+p)\Phi'$ becomes
\begin{equation*}
\frac{\kappa_N}{N-1}p=-(N-2)\frac{M}{r^N}+\lambda+\frac{1}{r}\left(1-\frac{2M}{r^{N-2}}-\lambda r^2\right)\frac{{\rm d}\Phi}{{\rm d}r},
\quad{\rm with}\quad
M(r)=\frac{\kappa_N}{N-1}\int_0^r\rho(\bar{r})\bar{r}^{N-1}{\rm d}\bar{r}.
\end{equation*}

Now, we consider taking derivative with respect to $r$ of the above equation,
\begin{equation}\label{appeq:buch1}
\frac{\kappa_N}{N-1}\frac{{\rm d}p}{{\rm d}r}=-(N-2)\frac{\rm d}{{\rm d}r}\left(\frac{M}{r^N}\right)+\frac{\rm d}{{\rm d}r}\left[\frac{1}{r}\left(1-\frac{2M}{r^{N-2}}-\lambda r^2\right)\frac{{\rm d}\Phi}{{\rm d}r}\right].
\end{equation}

In addition, note that, if ${\rm d}\rho/{\rm d}r\leq0$, the first term on the RHS of Eq.~(\ref{appeq:buch1})
\begin{equation}\label{appeq:buch2}
\frac{\rm d}{{\rm d}r}\left(\frac{M}{r^N}\right)=\left(\frac{{\rm d}M}{{\rm d}r}\right)\frac{1}{r^N}-\frac{NM}{r^{N+1}}=\frac{\kappa_N}{(N-1)r}\left(\rho-\bar{\rho}\right)<0,
\end{equation}
where $\bar{\rho}$ is defined through the mean value theorem for $0\leq r_0\leq r$,
\[
M(r)=\frac{\kappa_N}{N(N-1)}\rho(r_0)r^N\equiv\frac{\kappa_N}{N(N-1)}\bar{\rho}(r)r^N
\]
and $\bar{\rho}(r)\geq\rho(r)$ if ${\rm d}\rho/{\rm d}r\leq0$. Furthermore, the second term on the RHS of Eq.~(\ref{appeq:buch1}) can be written as
\[
\frac{\rm d}{{\rm d}r}\left[\left(\frac{\Phi'}{r}e^\Phi\sqrt{1-\frac{2M}{r^{N-2}}-\lambda r^2}\right)\left(e^{-\Phi}\sqrt{1-\frac{2M}{r^{N-2}}-\lambda r^2}\right)\right]
=\frac{\rm d}{{\rm d}r}\left[\left(\frac{\Phi'}{r}e^\Phi e^{-\Lambda}\right)\left(e^{-\Phi}e^{-\Lambda}\right)\right]
\]
\begin{equation}\label{appeq:buch3}
=e^{-\Phi}e^{-\Lambda}\frac{\rm d}{{\rm d}r}\left(\frac{\Phi'}{r}e^\Phi e^{-\Lambda}\right)
+\frac{\Phi'}{r}e^\Phi e^{-\Lambda}\frac{\rm d}{{\rm d}r}\left(e^{-\Phi}e^{-\Lambda}\right)
\quad
{\rm with}
\quad
e^{2\Lambda}=\left[1-\frac{2M}{r^{N-2}}-\lambda r^2\right]^{-1}
\end{equation}
and
\[
\frac{\rm d}{{\rm d}r}\left(e^{-\Lambda}\right)=-e^\Lambda\left[\frac{\rm d}{{\rm d}r}\left(\frac{M}{r^{N-2}}\right)+\lambda r\right].
\]

Therefore,
\[
e^{-\Lambda}\frac{\rm d}{{\rm d}r}\left(e^{-\Phi}e^{-\Lambda}\right)
=e^{-\Lambda}\left[-e^{-\Phi}e^\Lambda\left(\frac{\rm d}{{\rm d}r}\left(\frac{M}{r^{N-2}}\right)+\lambda r\right)-e^{-\Phi}\Phi' e^{-\Lambda}\right]
=-e^{-\Phi}\left\{\left[\frac{\rm d}{{\rm d}r}\left(\frac{M}{r^{N-2}}\right)+\lambda r\right]+\Phi' e^{-2\Lambda}\right\}
\]
\begin{equation}\label{appeq:buch4}
=-e^{-\Phi}\left\{\left[\left(\frac{\kappa_N\rho}{N-1}-\frac{(N-2)M}{r^N}+\lambda\right) r\right]+\Phi' e^{-2\Lambda}\right\}
=-e^{-\Phi}\left[\frac{\kappa_N(\rho+p)}{N-1}\right]r,
\end{equation}
where we have used the TOV equation in the last equality
\[
-\frac{(N-2)M}{r^N}+\lambda=\frac{\kappa_N}{N-1}p-\frac{1}{r}e^{-2\Lambda}\Phi' .
\]

Combining Eqs.~(\ref{appeq:buch1}),(\ref{appeq:buch2}),(\ref{appeq:buch3}),(\ref{appeq:buch4}), and the conservation law $p'=-(\rho+p)\Phi'$, it turns out
\begin{equation}\label{appeq:buch5}
(N-2)e^{\Phi+\Lambda}\frac{\rm d}{{\rm d}r}\left(\frac{M}{r^N}\right)=\frac{\rm d}{{\rm d}r}\left(\frac{\Phi'}{r}e^\Phi e^{-\Lambda}\right)\leq0.
\end{equation}

Integration of Eq.~(\ref{appeq:buch5}) from $r$ to $R$ gives
\[
\frac{\Phi'(R)}{R}e^{\Phi(R)-\Lambda(R)}-\frac{\Phi'(r)}{r}e^{\Phi(r)-\Lambda(r)}\leq0,
\]
also note that, in vacuum,
\[
e^{\Phi(r)}=\sqrt{1-\frac{2\mathcal{M}}{r^{N-2}}-\lambda r^2}=e^{-\Lambda(r)}
\quad{\rm for}\quad
r\geq R,
\]
so
\[
\Phi'(R)e^{\Phi(R)}=\frac{\rm d}{{\rm d}r}\left(e^{\Phi(r)}\right)\bigg|_{r=R}
=\frac{(N-2)\frac{\mathcal{M}}{R^{N-1}}-\lambda R}{\sqrt{1-\frac{2\mathcal{M}}{R^{N-2}}-\lambda R^2}}
=\left[(N-2)\frac{\mathcal{M}}{R^{N-1}}-\lambda R\right]e^{\Lambda(R)}
\]
\[
\Rightarrow
\frac{\Phi'(R)}{R}e^{\Phi(R)-\Lambda(R)}=\left[(N-2)\left(\frac{\mathcal{M}}{R^{N}}\right)-\lambda\right]
\Rightarrow
\frac{\Phi'(r)}{r}e^{\Phi(r)-\Lambda(r)}\geq\left[(N-2)\left(\frac{\mathcal{M}}{R^{N}}\right)-\lambda\right]
\]
or
\begin{equation}\label{appeq:buch6}
\Phi'(r)e^{\Phi(r)}\geq re^{\Lambda(r)}\left[(N-2)\left(\frac{\mathcal{M}}{R^{N}}\right)-\lambda\right].
\end{equation}

Integration of Eq.~(\ref{appeq:buch6}) from $0$ to $R$ yields
\begin{equation}\label{appeq:buch7}
e^{\Phi(R)}>e^{\Phi(R)}-e^{\Phi(0)}\geq
\left[(N-2)\left(\frac{\mathcal{M}}{R^{N}}\right)-\lambda\right]\int_0^R \frac{r{\rm d}r}{\sqrt{1-\frac{2M(r)}{r^{N-2}}-\lambda r^2}}
\end{equation}
and the non-increasing monotonicity of $\rho$ leads to the inequality
\[
M(r)=\frac{\kappa_N}{N(N-1)}r^N\bar{\rho}(r)=\left(\frac{r}{R}\right)^N\frac{\kappa_N}{N(N-1)}R^N\bar{\rho}(r)
\geq
\left(\frac{r}{R}\right)^N\frac{\kappa_N}{N(N-1)}R^N\bar{\rho}(R)=\left(\frac{r}{R}\right)^N \mathcal{M},
\]
hence
\begin{equation}\label{appeq:buch8}
\int_0^R \frac{r{\rm d}r}{\sqrt{1-\frac{2M(r)}{r^{N-2}}-\lambda r^2}}
\geq
\int_0^R \frac{r{\rm d}r}{\sqrt{1-\frac{2\mathcal{M}}{R^{N}}r^2-\lambda r^2}}
=\left[\frac{1-\sqrt{1-2\mathcal{M}/R^{N-2}-\lambda R^2}}{2\mathcal{M}/R^{N}+\lambda}\right]>0
\end{equation}
either for $2\mathcal{M}/R^N+\lambda>0$ or $<0$ if there is no horizon inside the sphere.
Consequently Eq.~(\ref{appeq:buch7}) and (\ref{appeq:buch8}) give rise to
\begin{equation*}
\sqrt{1-\frac{2\mathcal{M}}{R^{N-2}}-\lambda R^2}
=e^{\Phi(R)}>
\left[\left(\frac{N-2}{2}\right)\left(\frac{2\mathcal{M}}{R^{N-2}}\right)-\lambda R^2\right]
\left[\frac{1-\sqrt{1-2\mathcal{M}/R^{N-2}-\lambda R^2}}{2\mathcal{M}/R^{N-2}+\lambda R^2}\right]
\end{equation*}
or
\begin{equation}
\boxed{
\sqrt{1-2\mathcal{C}_N-a}>\left[(N-2)\mathcal{C}_N-a\right]\left[\frac{1-\sqrt{1-2\mathcal{C}_N-a}}{2\mathcal{C}_N+a}\right]
}
\end{equation}
with $\mathcal{C}_N\equiv\mathcal{M}/R^{N-2}>0$ and $a\equiv\lambda R^2$.
Note that if $2\mathcal{C}_N+a=0$, we have $1>N\mathcal{C}_N\times\frac{1}{2}\rightarrow 2\mathcal{C}_N=2\mathcal{M}/R^{N-2}=-\lambda R^2=-a< 4/N$. However, if $2\mathcal{C}_N+a\neq0$, one obtains
\[
N\mathcal{C}_N\sqrt{1-2\mathcal{C}_N-a}\gtrless\left[(N-2)\mathcal{C}_N-a\right]
\quad{\rm if}\quad
2\mathcal{C}_N+a\gtrless0.
\]

Both lead to the same inequality:
\begin{equation}
\mathcal{C}_N^2-\frac{2}{N}\left(\frac{N-1}{N}\right)\mathcal{C}_N+\frac{a}{N^2}<0
\end{equation}
and hence
\begin{equation}
\boxed{
\frac{N-1}{N^2}\left(1-\sqrt{1-\frac{N^2 a}{(N-1)^2}}\right)
<
\mathcal{C}_N
<
\frac{N-1}{N^2}\left(1+\sqrt{1-\frac{N^2 a}{(N-1)^2}}\right).
}
\end{equation}

\section{\label{app:n2tables}Various tables for $N=2$}
\setlength{\tabcolsep}{30pt} 
\renewcommand{\arraystretch}{1} 
\begin{table}[!ht]
  \caption{End points of $\lambda R^2$ for $N=2$ with various fluid mass $0<\mathcal{C}_2=\mathcal{M}\leq0.5$ at causal limit $v_c=1$.}
 \resizebox{\columnwidth}{!}{%
 \begin{tabular}{l c c c}
  \hline\hline
  $\mathcal{C}_2=\mathcal{M}$  & $\lambda R^2$ & $\langle\gamma\rangle$ & $\gamma_{{\rm cr}}$ \\ [0.5ex] 
  \hline\hline
$0.1$  & $-1.08681$ & $1.81892$ & $0.289458$ \\
$0.2$  & $-0.91554$ & $1.81892$ & $0.312447$ \\
$0.3$  & $-0.73198$ & $1.81892$ & $0.347248$ \\
$0.4$  & $-0.52604$ & $1.81893$ & $0.410982$ \\
$0.5$  & $-0.25$ & $1.81893$ & $0.636364$ \\
[0.5ex]
  \hline\hline
 \end{tabular}
 }
 \label{tab:III}
\end{table}

\setlength{\tabcolsep}{15pt} 
\renewcommand{\arraystretch}{1} 
\begin{table}[!ht]
  \caption{The lower and upper causal limits ($v_c=1$) of $\lambda R^2$ with various fluid mass $\mathcal{C}_2=\mathcal{M}>0.5$. The two causal limits become degenerate at $\mathcal{M}=0.5208\bar{3}$.}
 \resizebox{\columnwidth}{!}{%
 \begin{tabular}{l c c c}
  \hline\hline
  $\mathcal{C}_2=\mathcal{M}$  & $\lambda R^2$ (Lower/Upper) & $\langle\gamma\rangle$ (Lower/Upper) & $\gamma_{{\rm cr}}$ (Lower/Upper) \\ [0.5ex] 
  \hline\hline
$0.501$  & $-0.245968$~/~$-0.0020368908$ & $1.81892$~/~$1.81893$ & $0.643072$~/~$44.9933$\\
$0.504$  & $-0.233361$~/~$-0.00863897$ & $1.81893$~/~$1.81892$ & $0.665408$~/~$10.8801$ \\
$0.508$  & $-0.215107$~/~$-0.0188929$ & $1.81892$~/~$1.81892$ & $0.702112$~/~$5.16155$ \\
$0.512$  & $-0.194394$~/~$-0.0316059$ & $1.81893$~/~$1.81893$ & $0.751605$~/~$3.21809$ \\
$0.516$  & $-0.169208$~/~$-0.048792$ & $1.81893$~/~$1.81893$ & $0.827182$~/~$2.19555$ \\
$0.520$  & $-0.13$~/~$-0.08$ & $1.81893$~/~$1.81893$ & $1.00000$~/~$1.45455$ \\
$0.5208$  & $-0.1092$~/~$-0.0992$ & $1.81893$~/~$1.81893$ & $1.13986$~/~$1.22727$ \\
\hline
$0.5208\bar{3}$ & $-0.1041\bar{6}$ & $1.81893$ & $1.18182$ \\
[0.5ex]
  \hline\hline
 \end{tabular}
 }
 \label{tab:IV}
\end{table}

\setlength{\tabcolsep}{25pt} 
\renewcommand{\arraystretch}{1} 
\begin{table}[!ht]
  \caption{The minimal $v_c$ and the corresponding $\lambda R^2$ with various fluid mass $\mathcal{C}_2=\mathcal{M}>0.5$. $\mathcal{M}\simeq0.5208$ is the upper bound without violation of the causal limit $v_c=1$.}
 \resizebox{\columnwidth}{!}{%
 \begin{tabular}{l c c c c}
  \hline\hline
  $\mathcal{C}_2=\mathcal{M}$  & $\lambda R^2$ & $v_c=v(0)$ & $\langle\gamma\rangle$ & $\gamma_{{\rm cr}}$ \\ [0.5ex] 
  \hline\hline
$0.50001$  & $-0.0000401797$ & $0.134047$ & $1.99701$ & $1.00447$ \\
$0.5001$  & $-0.000405737$ & $0.239529$ & $1.99045$ & $1.01407$ \\
$0.501$  & $-0.00418706$ & $0.432509$ & $1.96890$ & $1.04396$ \\
$0.504$  & $-0.0175648$ & $0.625459$ & $1.93483$ & $1.08603$ \\
$0.508$  & $-0.0365920$ & $0.757625$ & $1.90366$ & $1.11908$ \\
$0.512$  & $-0.0566768$ & $0.850335$ & $1.87713$ & $1.14315$ \\
$0.516$  & $-0.0776784$ & $0.924608$ & $1.85193$ & $1.16248$ \\
$0.520$  & $-0.0995169$ & $0.987848$ & $1.82538$ & $1.17873$ \\
$0.5208$  & $-0.10398$ & $0.999520$ & $1.81921$ & $1.18170$ \\
\hline
$0.5208\bar{3}$ & $-0.1041\bar{6}$ & $1$ & $1.81893$ & $1.18182$ \\
[0.5ex]
  \hline\hline
 \end{tabular}
 }
 \label{tab:V}
\end{table}

\setlength{\tabcolsep}{35pt} 
\renewcommand{\arraystretch}{1} 
\begin{table}[!ht]
  \caption{The critical $\lambda R^2$ with various fluid mass $\mathcal{C}_2=\mathcal{M}>0.5$ under causal range $v_c\leq1$. $\mathcal{M}\simeq0.518$ is the upper bound at which the instability can be triggered without violation of causal limit $v_c=1$.}
 \resizebox{\columnwidth}{!}{%
 \begin{tabular}{l c c c }
  \hline\hline
  $\mathcal{C}_2=\mathcal{M}$  & $\lambda R^2$ & $v_c=v(0)$ & $\langle\gamma\rangle=\gamma_{{\rm cr}}$ \\ [0.5ex] 
  \hline\hline
$0.501$  & $-0.0027978$ & $0.461740$ & $1.96455$ \\
$0.504$  & $-0.0117926$ & $0.663440$ & $1.92657$ \\
$0.508$  & $-0.0247187$ & $0.799338$ & $1.89227$ \\
$0.512$  & $-0.0385385$ & $0.893364$ & $1.86307$ \\
$0.516$  & $-0.0532156$ & $0.967696$ & $1.83468$ \\
$0.518$  & $-0.0609066$ & $0.999979$ & $1.81894$ \\
\hline
$0.5180013$  & $-0.0609116$ & $1$ & $1.81893$ \\
[0.5ex]
  \hline\hline
 \end{tabular}
 }
 \label{tab:VI}
\end{table}

\section*{Data Availability Statement}
All data generated or analyzed during this study are included in this article.

\bibliographystyle{JHEP} 
\bibliography{dibh}

\end{document}